\documentclass[runningheads]{svmult}
\usepackage{makeidx}   
\usepackage{graphicx}  
\usepackage{subeqnar}  
\usepackage{multicol}  
\usepackage{physprbb}  
\makeindex             

\begin{document}

\title*{Imaging the secondary stars in CVs}
\toctitle{Imaging the secondary stars in CVs}
\titlerunning{Imaging the secondary stars in CVs}
\author{V. S. Dhillon \and C. A. Watson}
\authorrunning{V. S. Dhillon \and C. A. Watson}
\institute{Department of Physics and Astronomy, University of Sheffield,
Sheffield S3 7RH, UK}
\maketitle              

\begin{abstract}
The secondary, Roche-lobe filling stars in cataclysmic variables (CVs)
are key to our understanding of the origin, evolution and behaviour of
this class of interacting binary. We review the basic properties of
the  secondary stars in CVs and the observational and analysis methods
required to detect them. We then describe the various
astro-tomographic  techniques which can be used to map the surface
intensity distribution  of the secondary star, culminating in a
detailed explanation of Roche tomography. We conclude with a summary
of the most important results  obtained to date and future prospects.
\end{abstract}

\section{Introduction}

\begin{figure}[t]
\begin{center}
\includegraphics{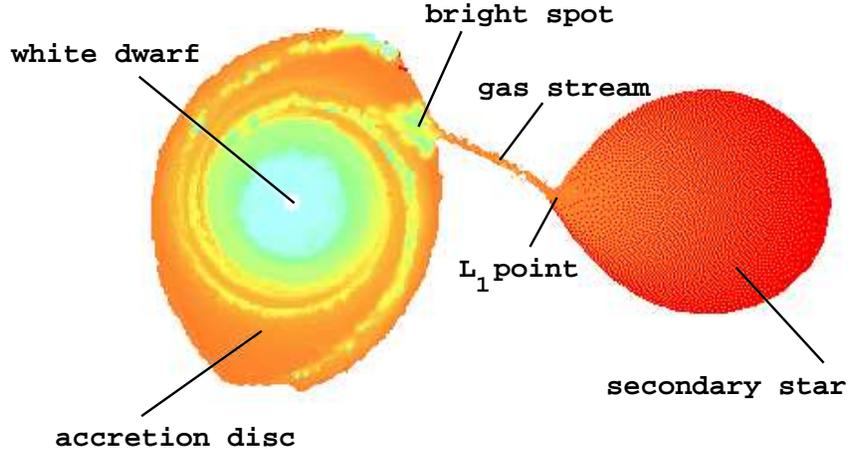}
\end{center}
\caption[]{A non-magnetic CV.}
\label{fig:cvcartoon}
\end{figure}

CVs are semi-detached binary stars consisting of a Roche-lobe filling
secondary star transferring mass to a white dwarf primary star via an
accretion disc or magnetically-channelled accretion flow.  The CVs are
classified into a number of different sub-types, including the  {\em
novae}, {\em recurrent novae}, {\em dwarf novae}, and {\em novalikes},
according to the nature of their cataclysmic (i.e. violent  but
non-destructive) outbursts. A further sub-division into {\em polars}
and {\em intermediate polars} is also made if a CV accretes via magnetic
field lines. Figure~\ref{fig:cvcartoon} depicts the five principal
components of a typical non-magnetic CV: the primary star, the
secondary star, the gas stream (formed by the transfer of material
from the secondary to the primary), the accretion disc and the bright
spot (formed by the collision between the gas stream and the edge of
the accretion disc). The distance between the stellar components is
approximately a solar radius and the orbital period is typically a few
hours. For a detailed review of these objects, see~\cite{warner95}.

\subsection{The nature of the secondary stars in CVs}

The spectral type and luminosity class of the secondary star in CVs
can  be estimated from basic theory, as follows. The mean density,
${\bar\rho}_2$, of the secondary star is given by
\begin{equation}
\label{eq:rho}
{\bar\rho}_2 = {M_2 \over {4\over 3}\pi R_2^3},
\end{equation}
where $M_2$ and $R_2$ are the mass and volume radius (i.e. the  radius
of a sphere with the same volume as the Roche lobe) of the secondary
star, respectively. The volume radius of the Roche lobe can be
approximated by
\begin{equation}
{{R_2} \over a} = 0.47 \biggl({q \over {1+q}}\biggr)^{1/3}
\label{eq:volrad}
\end{equation}
\cite{smith98a}, where $q=M_2/M_1$ and $a$ is the distance between the
centres of mass of the  binary components. Newton's generalisation of
Kepler's third law can be  written as
\begin{equation}
{{4\pi^2 a^3} \over GP^2} = M_1 + M_2 = M_2 \biggl({{1+q} \over q}\biggr),
\label{eq:kepler3}
\end{equation}
where $P$ is the orbital period of the binary. Combining
equations~\ref{eq:rho},~\ref{eq:volrad} and~\ref{eq:kepler3} gives the
mean density-orbital period relation,
\begin{equation}
{\bar\rho}_2 = 105\, P^{-2} {\rm \ (h)} {\ \ \ \rm g\ cm}^{-3},
\end{equation}
which is accurate to $\sim$6 per cent~\cite{eggleton83} over the range
of mass ratios relevant to most CVs ($0.01<q<1$). Most CVs have
orbital periods of $1.25 {\rm \ h} < P < 9 {\rm \ h}$, resulting in
mean densities of $67  {\rm \ g\,cm}^{-3} < {\bar\rho}_2 < 1.3 {\rm \
g\,cm}^{-3} $. Such mean densities are typically found in M8V--G0V
stars~\cite{allen76} and hence the secondary stars in CVs should be 
similar to M, K or G main-sequence dwarfs. With a few caveats, this 
prediction is largely confirmed by 
observation~\cite{smith98a},\cite{beuermann98}.

\subsection{Why image the secondary stars in CVs?}

Even though we have just shown that most CV secondary stars are
similar to lower main-sequence stars in their gross properties, it is
not clear that they should share the same detailed properties. This is
because CV  secondaries are subject to a number of extreme
environmental factors to  which isolated stars are not. Specifically,
CV secondaries are:
\begin{enumerate}
\item 
situated $\sim 1R_\odot$ from a hot, irradiating source (see~\cite{smith95});
\item
rapidly rotating ($v_{rot} \sim 100$~km\,s$^{-1}$);
\item
Roche-lobe shaped;
\item
losing mass at a rate of $\sim10^{-8} - 10^{-11}M_\odot$\,yr$^{-1}$;
\item 
survivors of a common-envelope phase during which they existed within
the atmosphere of a giant star, and
\item 
exposed to nova outbursts every $\sim 10^4$\,yr. 
\end{enumerate}

In order to study the impact of some of these environmental factors on
the detailed properties of CV secondary stars, surface images are
required. Direct imaging is impossible, however, as typical CV
secondary stars have radii of 400\,000 km and distances of 200
parsecs, which means that to detect a feature covering 20 per cent of
the star's surface requires a resolution of approximately 1
micro-arcsecond, 10\,000 times greater than the diffraction-limited
resolution of the world's largest telescopes. Astro-tomography is
hence a necessity when studying surface structure on CV secondaries.

Obtaining surface images of CV secondaries has much wider
implications, however, than just providing information on the detailed
properties of these stars. For example, a knowledge of the irradiation
pattern on the inner hemisphere of the secondary star in CVs is
essential if one is to calculate stellar masses accurate enough to
test binary star evolution models  (see section~3.3.4). Furthermore,
the irradiation pattern provides information on the geometry of the
accreting structures around  the white dwarf (see section~3.3.4).
Perhaps even more importantly, surface images of CV secondaries can be
used to study the solar-stellar connection.  It is well known that
magnetic activity in isolated lower-main sequence stars increases with
decreasing  rotation period (e.g.~\cite{rutten87}). The most rapidly
rotating isolated stars of this type have rotation periods of $\sim 8$
hours, much slower  than the synchronously rotating secondary stars
found in most CVs. One would therefore expect CVs to show even higher
levels of magnetic activity. There  is a great deal of indirect
evidence for magnetic activity in CVs -- magnetic activity cycles have
been invoked to explain variations in the orbital periods, mean
brightnesses and mean intervals between outbursts in CVs
(see~\cite{warner95}). The magnetic field of the secondary star is
also believed to play a crucial role in angular momentum loss via
magnetic braking in longer-period CVs, enabling CVs to transfer mass
and evolve to shorter periods. One of the observable consequences of
magnetic activity are star-spots, and their number, size, distribution
and variability, as deduced from astro-tomography of CV secondaries,
would provide critical tests of stellar dynamo models in a hitherto
untested period regime.

\section{Detecting the secondary stars in CVs}

Detecting spectral features from the secondary stars in CVs is not
easy.  There are two main reasons for this. First, CVs are typically
hundreds of  parsecs distant, rendering the lower main-sequence
secondary very faint.  Second, the spectra of CVs are usually
dominated by the accretion disc and the resulting shot-noise
overwhelms the weak signal from the secondary star. The result is
that, in the 1998 study of Smith and Dhillon~~\cite{smith98a}, only 55
of the 318 CVs with measured orbital periods had  spectroscopically
identified secondary stars.

\begin{figure}[t]
\begin{center}
\includegraphics[width=0.485\textwidth]{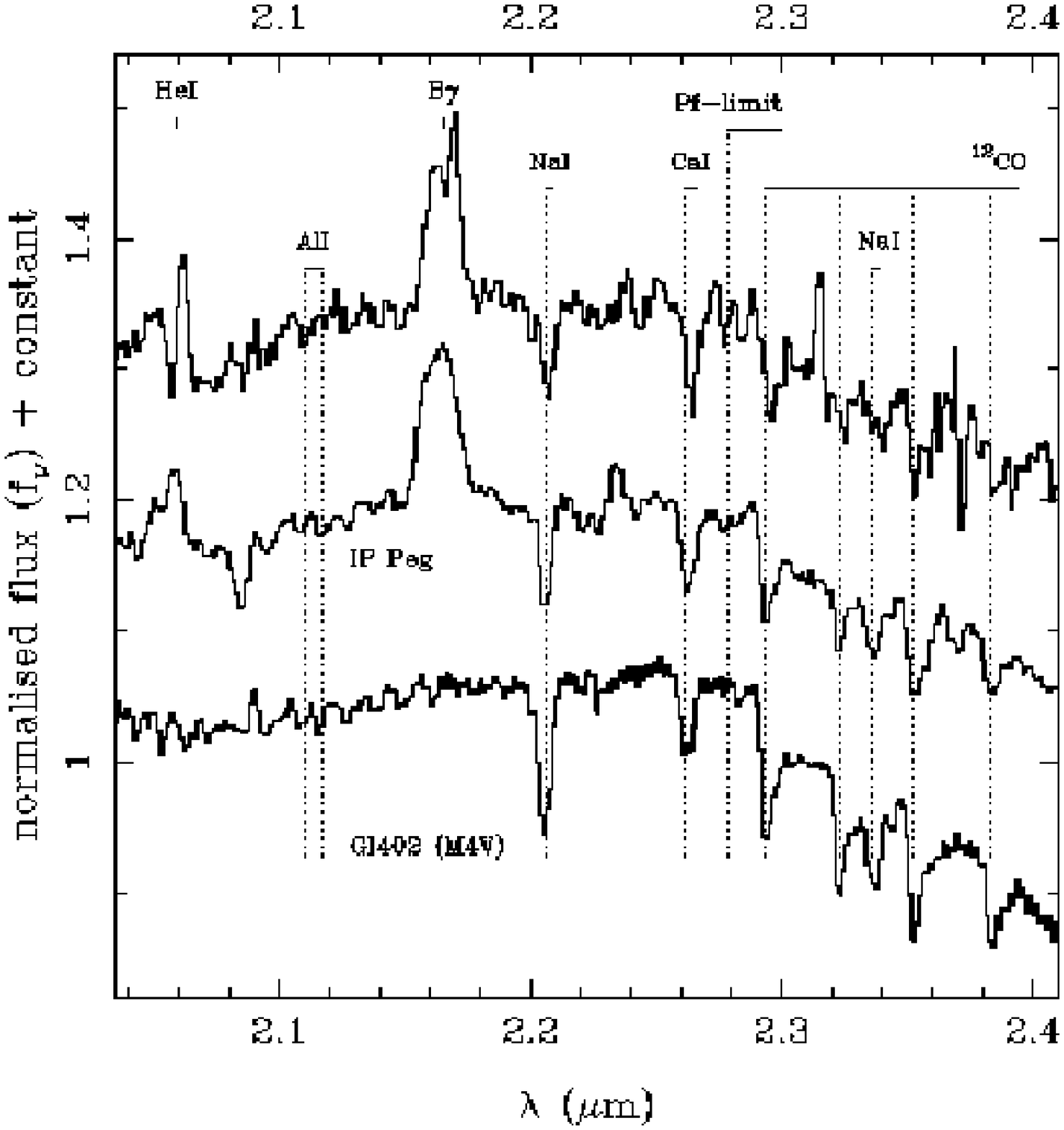}~\includegraphics[width=0.485\textwidth]{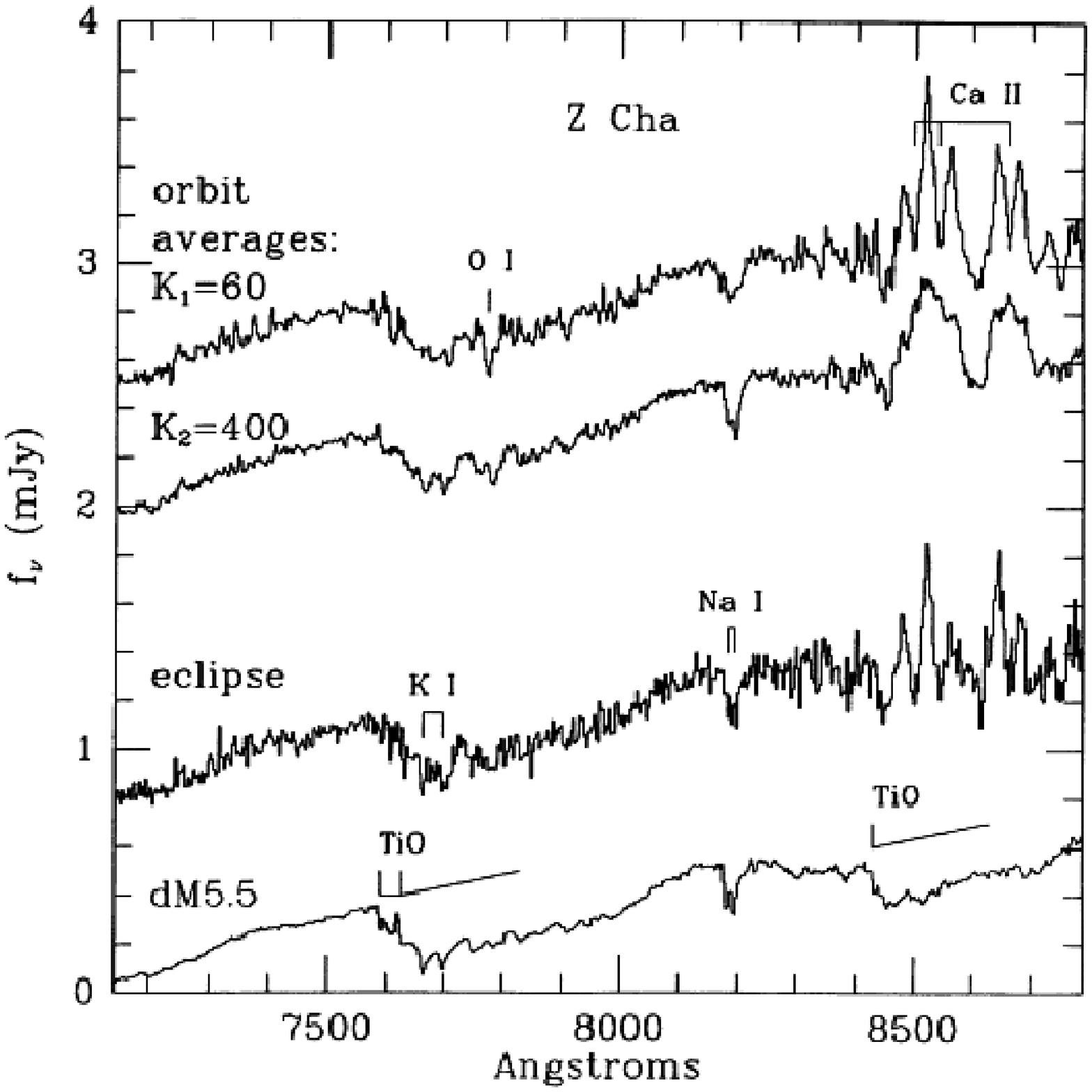}
\includegraphics[width=0.485\textwidth]{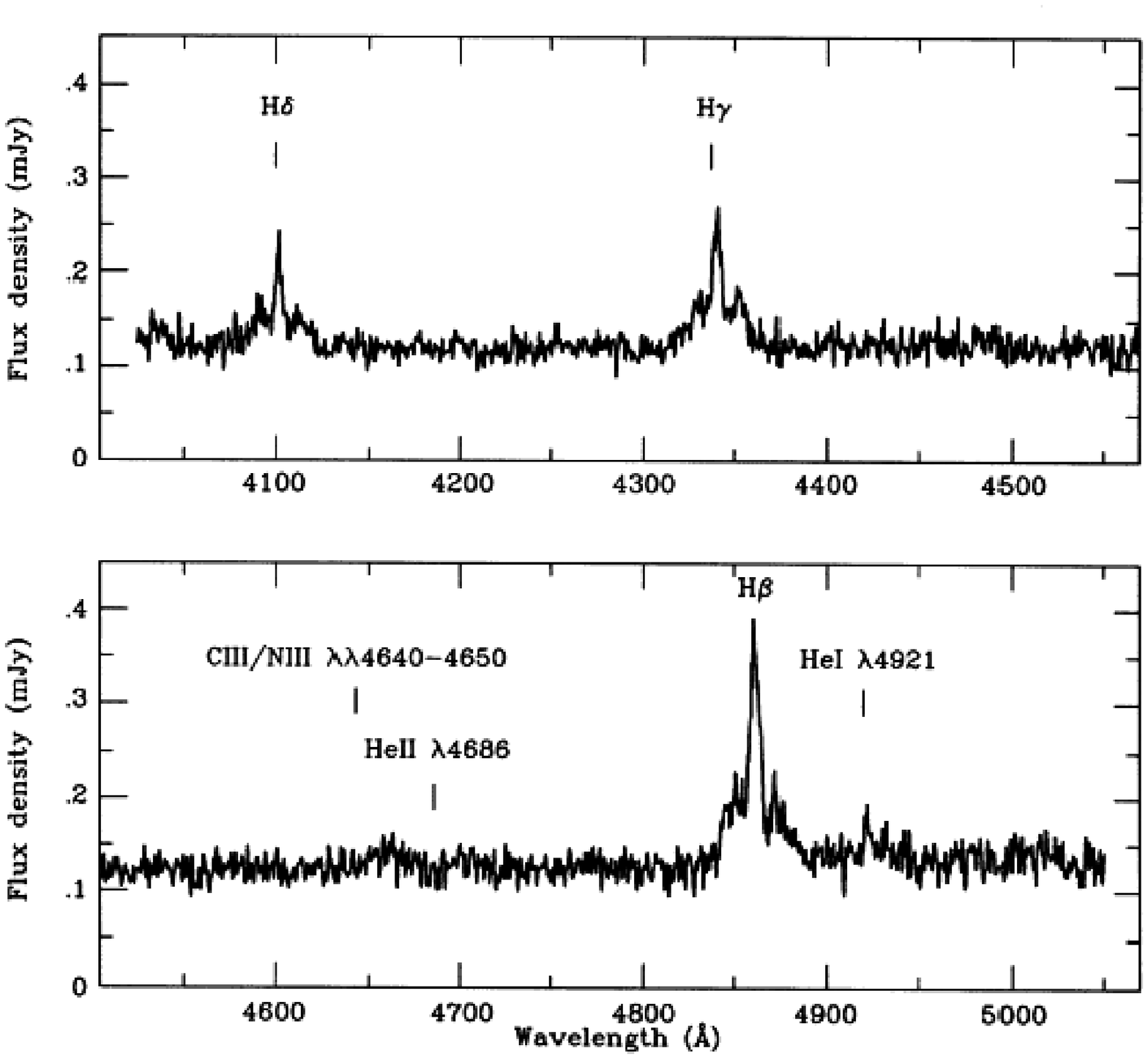}~\includegraphics[width=0.485\textwidth]{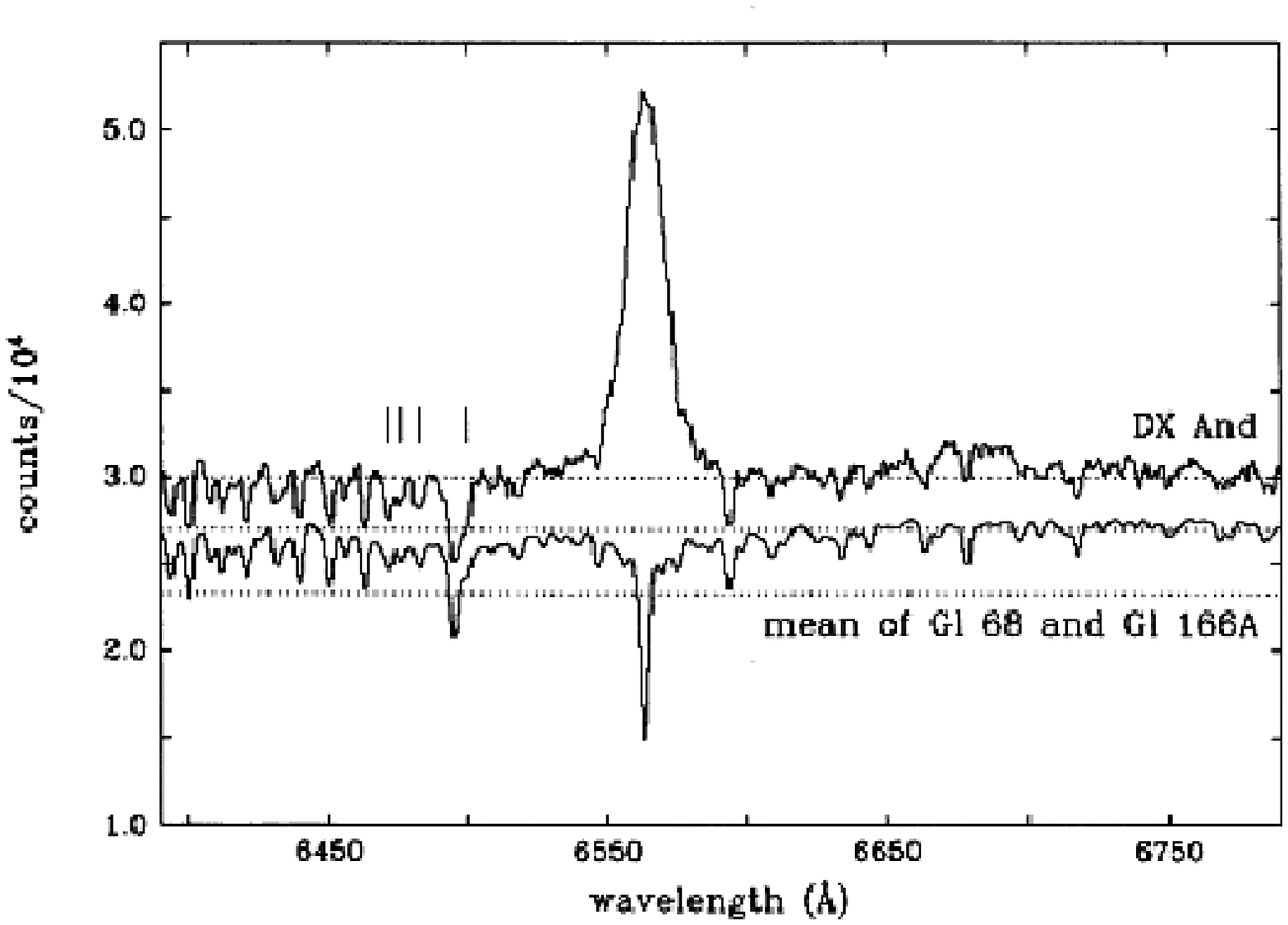}
\end{center}
\caption[]{Secondary star features in CV spectra. Clockwise from
top-left: IP~Peg~\protect\cite{littlefair00b}, Z~Cha~\protect\cite{wade88},
DX~And~\protect\cite{drew93} and DW~UMa~\protect\cite{dhillon94}.}
\label{fig:spectra}
\end{figure}

The best secondary star detection strategy depends very much on the 
orbital period of the CV, as this approximately determines the spectral type 
of the secondary via the empirical relation
\begin{equation}
\begin{array}{rrrll}
Sp(2)\,\,\, = & \,\,26.5 & - \  0.7 & \ P ({\rm h}), & P < 4\,{\rm h} \\ 
& \pm \,\, 0.7 &  \pm \,\, 0.2 & \\ 
& & & & \\ 
= & \,\,33.2 & - \ 2.5 & \ P ({\rm h}), & P \geq 4\,{\rm h} \\
& \pm \,\, 3.1 & \pm \,\, 0.5 & \\
\end{array}
\label{eq:deneal}
\end{equation}
\cite{smith98a}, where $Sp(2)$ is the spectral type of the secondary;
$Sp(2)=0$ represents  a spectral type of G0, $Sp(2)=10$ is K0 and
$Sp(2)=20$ is M0. Longer period CVs therefore have earlier
spectral-type secondaries, which generally contribute a greater
fraction (typically $>$75 per cent) of the total optical/infrared
light than the secondary stars found in shorter period CVs, which
usually only contribute $\sim$10--30 per cent~\cite{dhillon00}.
Equation~\ref{eq:deneal} can then be used in conjunction with a
black-body approximation to the wavelength, $\lambda_{\rm max}$, of
the peak flux, $f_{\nu}$, for a star of effective temperature
$T_{\rm eff}$,
\begin{equation}
\lambda_{\rm max} = 5100 / T_{\rm eff} \ \ \ \mu{\rm m}, 
\end{equation}
to obtain a crude idea of the optimum observation wavelength required
to detect the secondary star in a CV of known orbital period. With
$T_{\rm eff}$ ranging from $\sim$6000--2000~K for the G--M dwarf
secondary stars found in most CVs, $\lambda_{\rm max}$ ranges from
$\sim$0.8--2.5 $\mu$m, i.e. the optimum wavelength always lies in the
optical and near-infrared. In practice, when observing the secondary
stars in longer-period CVs it is  generally best to observe the
numerous neutral metal absorption lines in the  R-band around
H$\alpha$ (e.g.~\cite{drew93};  bottom-right,
figure~\ref{fig:spectra}), whereas short and intermediate-period
secondaries are best observed via the TiO molecular bands and Na\,I
absorption doublet in the I-band (e.g.~\cite{wade88}; top-right,
figure~\ref{fig:spectra}). If optical spectroscopy fails to detect the
secondary star in a CV, as is often the case in the shortest-period
dwarf novae and the novalikes (which have very bright discs), it is
possible to use near-infrared spectroscopy in the
J-band~\cite{littlefair00a} and K-band (e.g.~\cite{dhillon00}); the
brighter background in the near-infrared is offset to some extent by
the greater line  flux from the secondary star at these wavelengths
(e.g.~\cite{littlefair00b}; top-left, figure~\ref{fig:spectra}).  In
addition to absorption-line features, emission-line features from the
secondary star are sometimes visible in CV spectra, especially in
dwarf novae during outburst and novalikes during low-states
(e.g.~\cite{dhillon94}; bottom-left, figure~\ref{fig:spectra}).  These
narrow, chromospheric emission lines originate on  the inner
hemisphere of the secondary star and are most probably due  to
irradiation from the primary and its associated accretion regions. The
emission lines are usually most prominent in the Balmer lines, but
have also  been observed in He\,I, He\,II and Mg\,II
(e.g.~\cite{harlaftis99}).

\subsection{Skew mapping}

If the secondary star is not visible in a single spectrum of a CV, one
might naturally think that co-adding additional spectra will increase
the signal-to-noise and hence increase the chances of a
detection. This is true, but only if the additional spectra are first
shifted to  correct for the orbital motion of the secondary star, as
otherwise the weak features will be smeared out. The problem is that
the orbital motion is not known in advance; the solution is to use a
technique known as {\em skew mapping}~\cite{smith93}.

The first step is to cross-correlate each spectrum to be co-added with
a template, usually the spectrum of a field dwarf of matched spectral
type, yielding a time-series of cross-correlation functions (CCFs). If
there is a strong correlation, the locus of the CCF peaks will trace
out a sinusoidal path in a `trailed spectrum' of CCFs, in which case
plotting the velocity of each of the CCF peaks versus orbital phase
allows one to define the secondary star orbit. More often, however,
the  CCFs are too noisy to enable well-defined peaks to be
measured. Instead, the trailed spectrum of CCFs is back-projected in
an identical manner to that employed in standard Doppler
tomography~\cite{marsh00} to produce  what is known as a {\em skew
map}. Any noisy peaks in the trailed spectrum of CCFs which lie along
the true sinusoidal path of the secondary star will re-inforce during
the back projection process, resulting in a  spot on the skew map at
($0, K_2$), where $K_2$ is the radial-velocity semi-amplitude of the
secondary star.

\begin{figure}[t]
\begin{center}
\rotatebox{270}{
\includegraphics[width=0.57\textwidth]{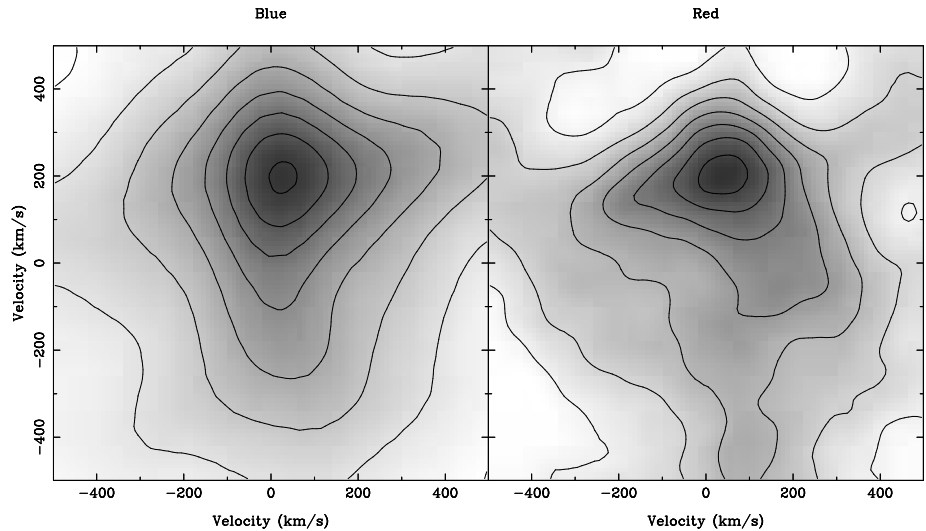}
}
\end{center}
\caption[]{Skew maps of the old nova BT~Mon~\protect\cite{smith98b}.}
\label{fig:skewmap}
\end{figure}

Figure~\ref{fig:skewmap} shows an example of the successful use of the
skew mapping technique applied to the old nova BT Mon~\cite{smith98b}.
The value of $K_2$ determined from the skew map (205~km\,s$^{-1}$) was
used to produce a co-added  spectrum which enabled the rotational
broadening and spectral type of the  secondary star to be accurately
determined. These system parameters were then used to calculate the
distance and the component masses of BT Mon, which provide fundamental
input to thermonuclear runaway models of nova outbursts. Although a
powerful technique which is invaluable in the detection of faint
secondary stars in CVs, skew mapping does not, however, provide
surface images. To do this,  other techniques are required, which we
shall turn to now.

\section{Mapping the secondary stars in CVs}

The secondary stars in CVs are three-dimensional objects. Time-series
of astronomical data, however, are effectively one-dimensional (in the
case of  light curves) or two-dimensional (in the case of
spectra). The problem of obtaining surface images of CV secondaries is
hence poorly constrained (see~\cite{rice89}), especially if one is
restricted to the so-called one-dimensional case. We will begin the
description of CV secondary mapping techniques by looking at the
one-dimensional techniques. These  include photometric light curve
fitting, but also include methods where spectral information is
parameterised in some way and the resulting values are then used to
obtain surface images; radial-velocity curve fitting, line-flux
fitting and line-width fitting all fall into this latter category. We will
then look at the special case of Doppler tomography, which uses
two-dimensional data to map two-dimensional structures. This means
that Doppler tomography is fully constrained, but it also means that
the secondary star is effectively compressed along the direction
defined by the rotation axis into only two dimensions.  Finally, we
will look at potentially the most powerful technique -- Roche
tomography -- which is very similar to the single-star mapping
techniques described elsewhere in this
volume~\cite{cameron00},\cite{donati00}  and which uses
two-dimensional data to construct three-dimensional surface images of
CV secondaries.

\subsection{One-dimensional techniques}
\label{sec:1d}

\subsection*{3.1.1~~~Radial-velocity curve fitting}

If the centre-of-light and centre-of-mass of the secondary are not
coincident,  the star's radial-velocity curve will be distorted in
some way from the pure sine wave which represents the motion of the
centre-of-mass. The observed radial-velocity curves of CV secondaries
suffer from this distortion, due to both geometrical effects caused by
the Roche-lobe shape and non-uniformities in the surface distribution
of the line strength due to, for example, irradiation.  Davey and
Smith~\cite{davey92},\cite{davey96} introduced an inversion technique
which exploits this effect and uses the asymmetries present in
observed radial-velocity curves to produce surface images of  CV
secondaries. Their `one-spot' model assumed a single region of heating
on the inner hemisphere of the secondary star which was allowed to
vary in strength, size and position, i.e. there were 3 free
parameters. The position of the spot, however, was restricted to vary
in only the longitudinal direction, and hence the resulting maps
assume symmetry about the orbital plane and are only two-dimensional.

\begin{figure}[t]
\begin{center}
\parbox{0.49\textwidth}{
\includegraphics[width=0.485\textwidth]{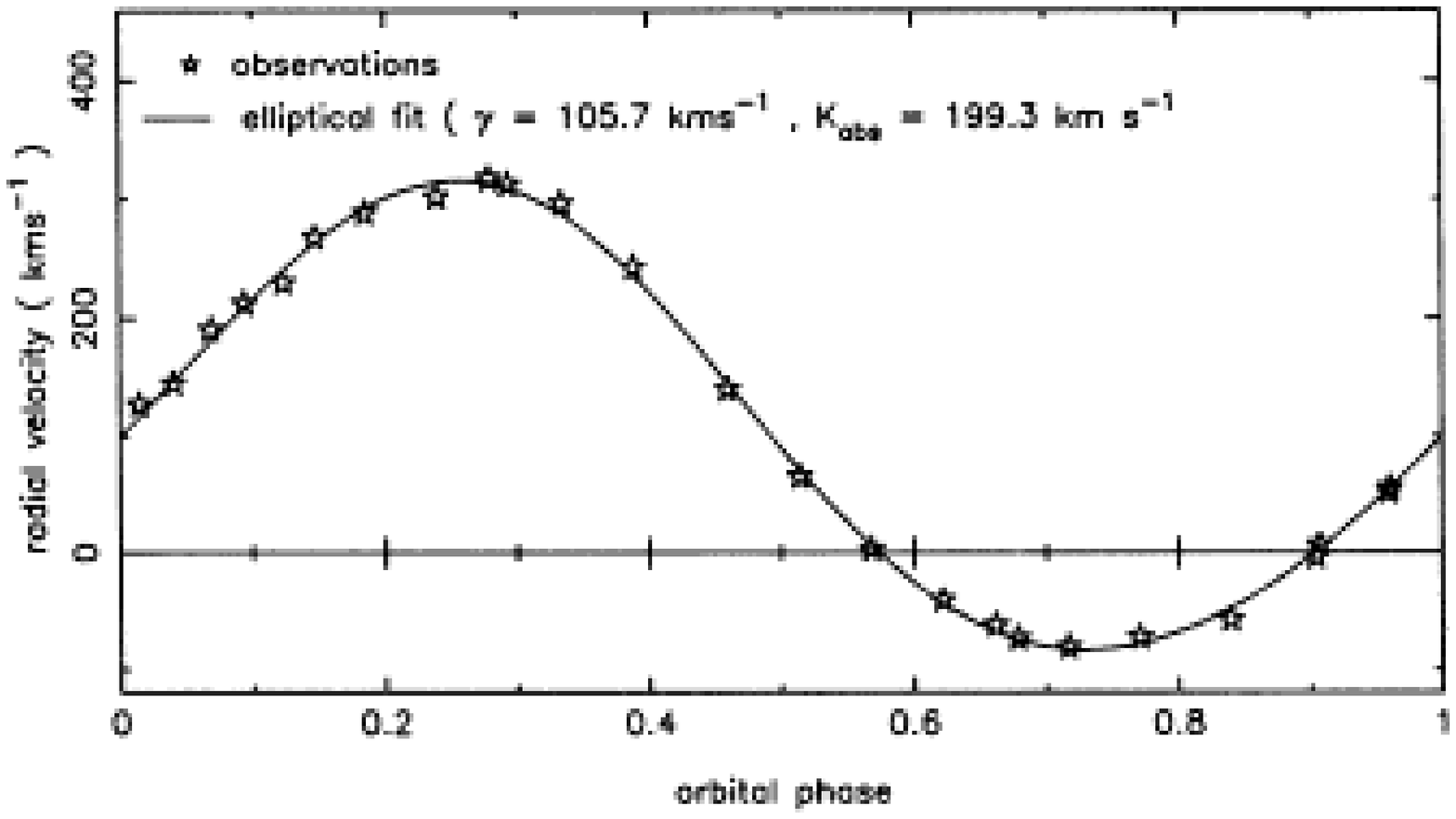}
\includegraphics[width=0.49\textwidth]{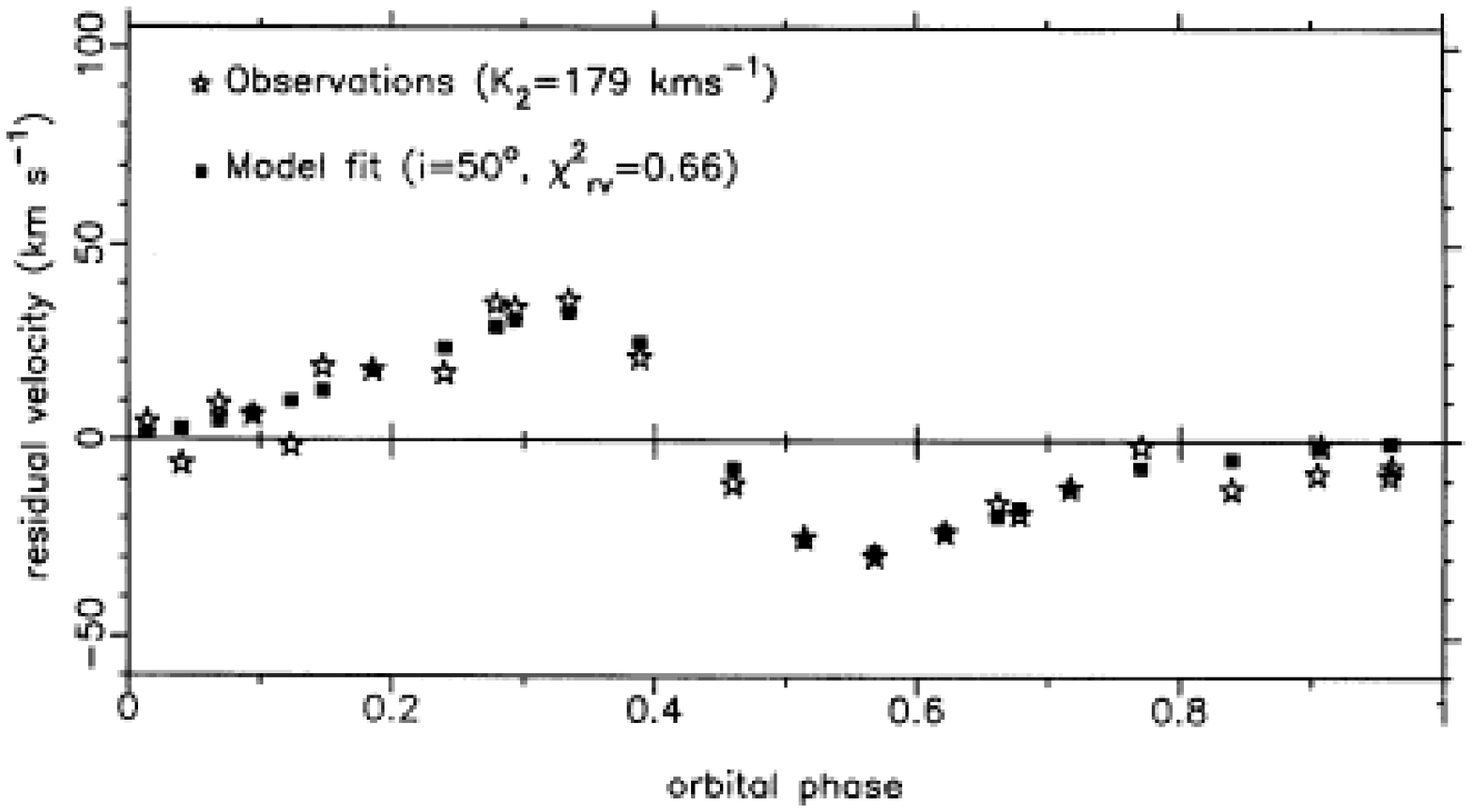}
}
\parbox{0.49\textwidth}{
\includegraphics[width=0.49\textwidth]{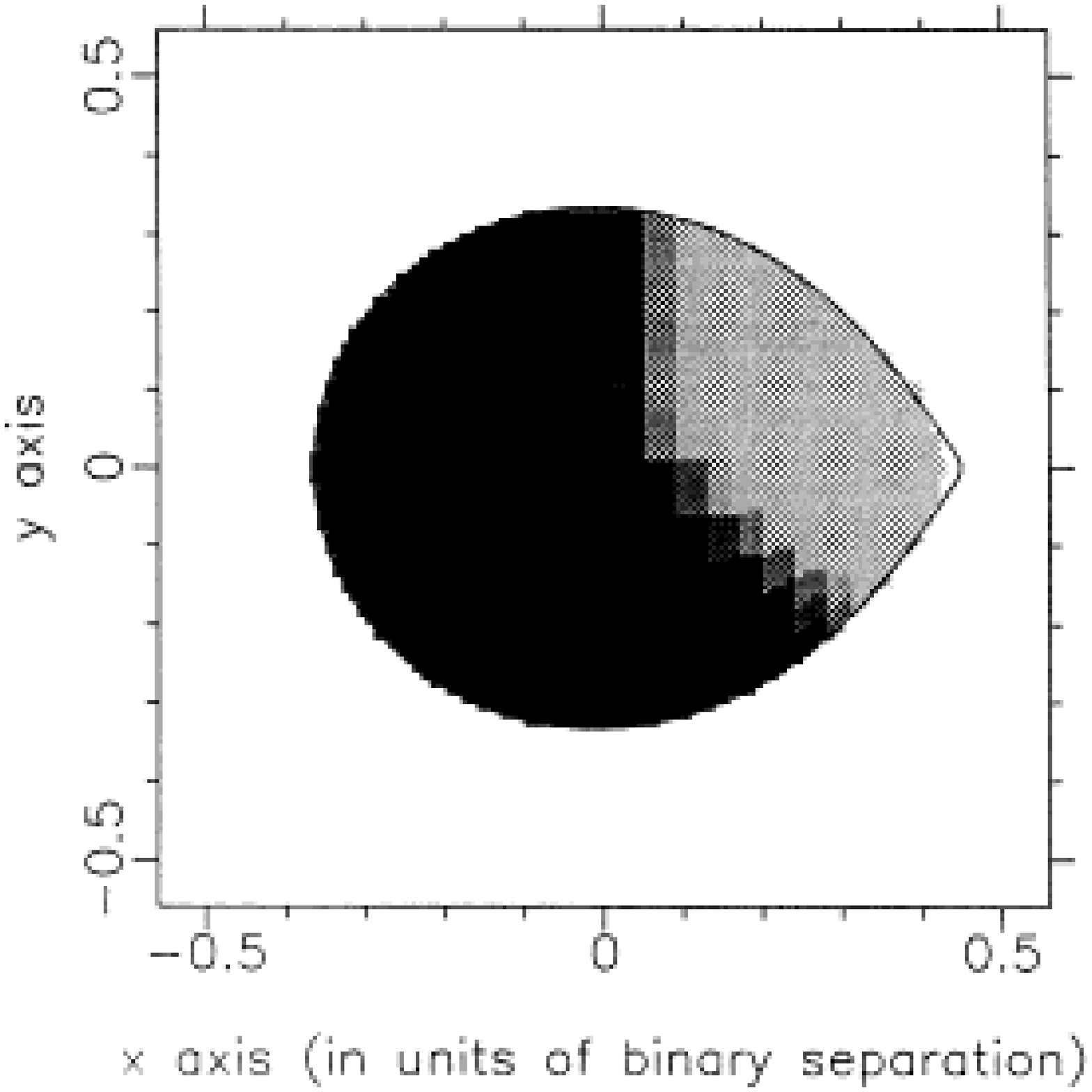}
}
\caption[]{Radial-velocity curve fitting of the polar AM~Her.
Upper-left: Radial-velocity curve of the Na\,I absorption doublet
around 8190\AA~\protect\cite{davey92}. Lower-left: Residual-velocity
curve, i.e. the motion of the centre-of-mass of the secondary has been
removed~\protect\cite{davey96}.  Right: Surface
map of AM~Her -- the light region is  where the
Na\,I doublet is weakest~\protect\cite{davey96}.}
\label{fig:davey}
\end{center}
\end{figure}

Davey and Smith have successfully applied their technique to four
dwarf novae and one
polar~\cite{davey92},\cite{smith95},\cite{davey96}.  The surface image
they derived of the polar AM~Her is shown in the right-hand panel of
figure~\ref{fig:davey}. It can be seen that there is a general
reduction in the line strength on the inner hemisphere of the
secondary, a result consistent with the effects of heating caused by
irradiation~\cite{brett93}. It can also be seen that the line
absorption is stronger on the leading  hemisphere (the lower edge of
the map in figure~\ref{fig:davey}) than it  is on the trailing
hemisphere (the upper edge of the map in figure~\ref{fig:davey}). This
asymmetry has been interpreted as due to the gas stream blocking the
radiation produced in the magnetic accretion column close to the white
dwarf, thereby shielding an area on the leading hemisphere of the
secondary star from irradiation. A similar asymmetry, but in the
opposite sense (i.e. the trailing hemisphere has stronger line
absorption than the leading hemisphere), has also been seen in the
dwarf novae mapped by Davey and Smith~\cite{smith95}. In these
objects, however, there is an accretion disc instead of an accretion
column.  Irradiation by the bright-spot is apparently insufficient to 
account for the observed asymmetry~\cite{davey92}, and so Davey and Smith
speculated that circulation currents induced by the irradiation spreads
the heating over a larger area, but preferentially towards the leading
hemisphere due to coriolis  effects. This idea was later confirmed by
SPH modelling~\cite{martin95}.

\subsection*{3.1.2~~~Light-curve fitting}

Light-curve fitting is now a well-established tool in CV research,
having found particular success in the study of accretion discs via
the eclipse  mapping method \cite{horne85},\cite{baptista00}.  The
brightness of the  secondary star in eclipse mapping is, however,
either completely ignored or  included in the fit as a single nuisance
parameter~\cite{rutten94a}. In order to obtain surface images of CV
secondaries via light-curve fitting, it is necessary to extend the
eclipse mapping method by constructing a secondary star grid of tiles
in addition to, or instead of, an accretion disc grid and fit the
whole  light curve rather than just the eclipse portion. Each element
in the grid is  assigned an intensity, which is weighted according to
its projected area and limb darkening at each orbital phase. A model
light curve is then derived by  summing the weighted intensities as a
function of orbital phase. By comparing the model light curve for some
grid brightness distribution with the observed light curve, the
element intensities can be iteratively adjusted until the observed
light curve is optimally fitted, usually using the maximum-entropy
approach described in section~3.3.1.

\begin{figure}[t]
\begin{center}
\rotatebox{0.5}{
\includegraphics[width=0.975\textwidth]{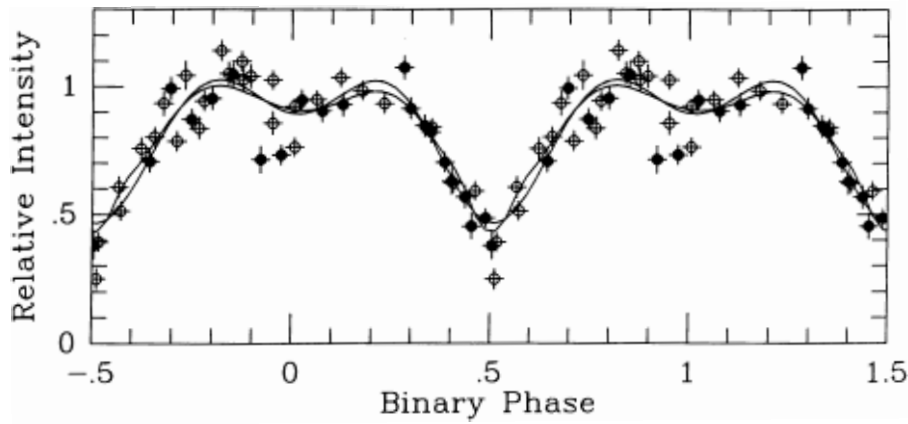}
}
\end{center}
\caption[]{Ellipsoidal variations in the dwarf nova 
Z~Cha~\protect\cite{wade88}.}
\label{fig:zcha}
\end{figure}

Such an approach has been adopted by Wade and Horne~\cite{wade88}, who
used a secondary star grid, but no accretion disc grid, to fit the TiO
band-flux light curve of the dwarf nova Z~Cha. The observed data is
represented by the points in  figure~\ref{fig:zcha} and their fits to
the data are represented by the solid lines. The observed data show a
double-humped variation, known as an {\em ellipsoidal
modulation}. This is due to the changing aspect of  the secondary's
Roche-lobe which presents the largest projected  area,  and hence
highest flux, at quadrature (phases 0.25 and 0.75) and the  smallest
project area, and hence lowest flux, during conjunction (phases  0 and
0.5).  The best fit to the data was obtained with a TiO distribution
which has a minimum around the $L_1$ point and rises  smoothly to a
level 3 times higher on the hemisphere facing away from the white
dwarf. This surface variation is consistent with the effects of
irradiation~\cite{brett93} and was used to correct the
radial-velocity  curves  for the effects of a mis-match between the
secondary's light centre and its centre-of-mass in order to  derive
accurate stellar masses.

\begin{figure}[t]
\begin{center}
\parbox{0.49\textwidth}{
\includegraphics[width=0.485\textwidth]{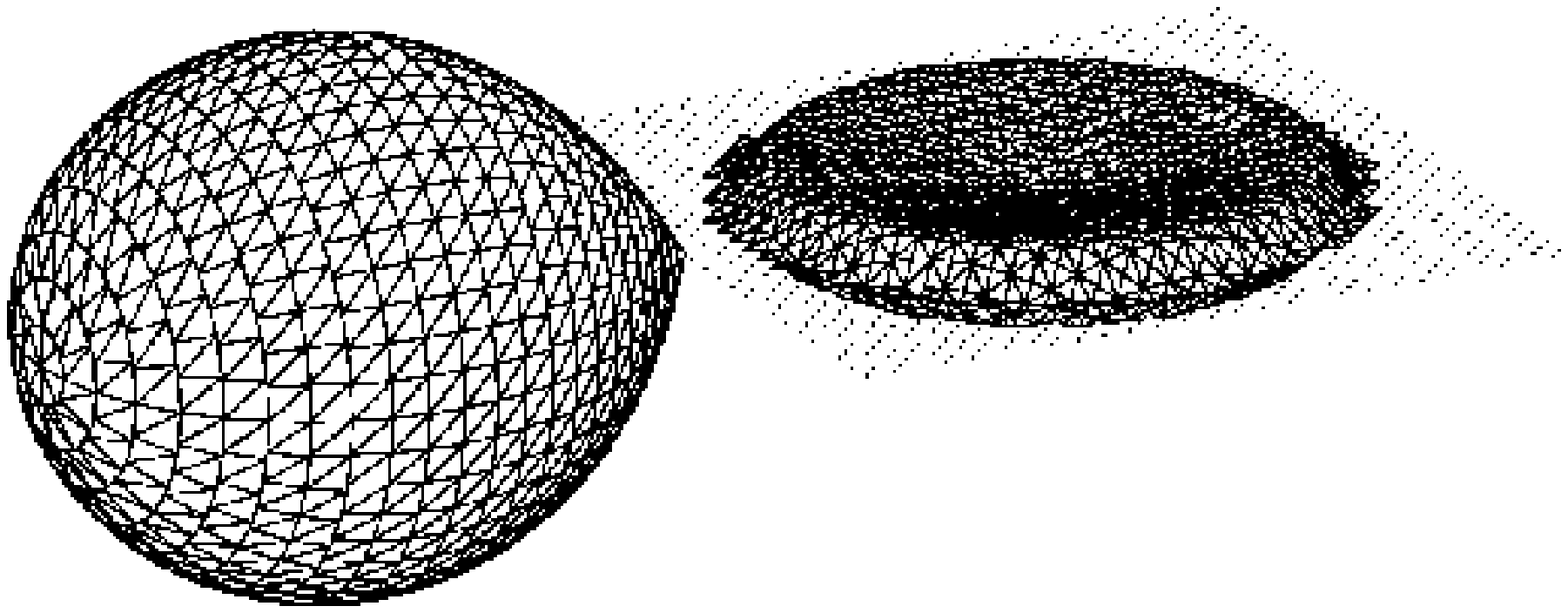}

\vspace*{0.25cm}

\includegraphics[width=0.49\textwidth]{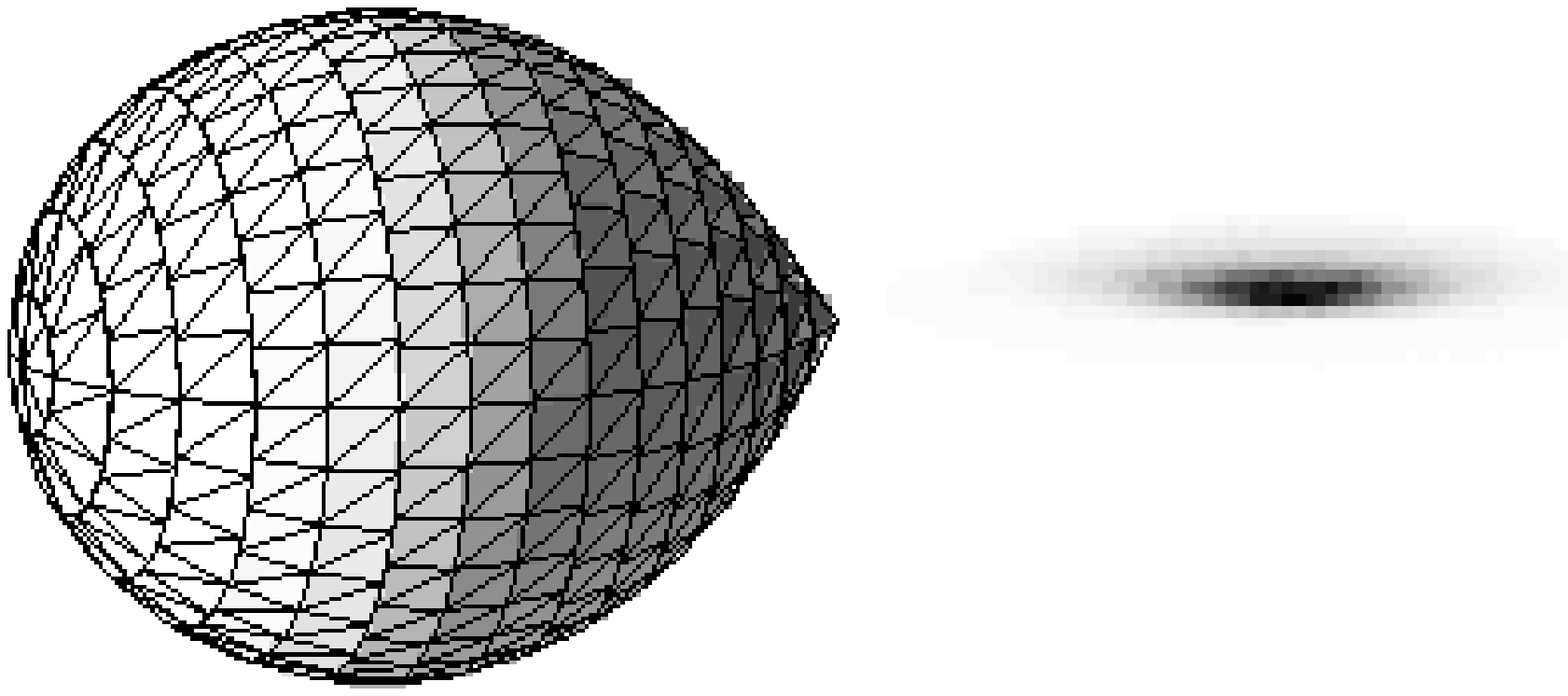}
}
\parbox{0.49\textwidth}{
\includegraphics[width=0.49\textwidth]{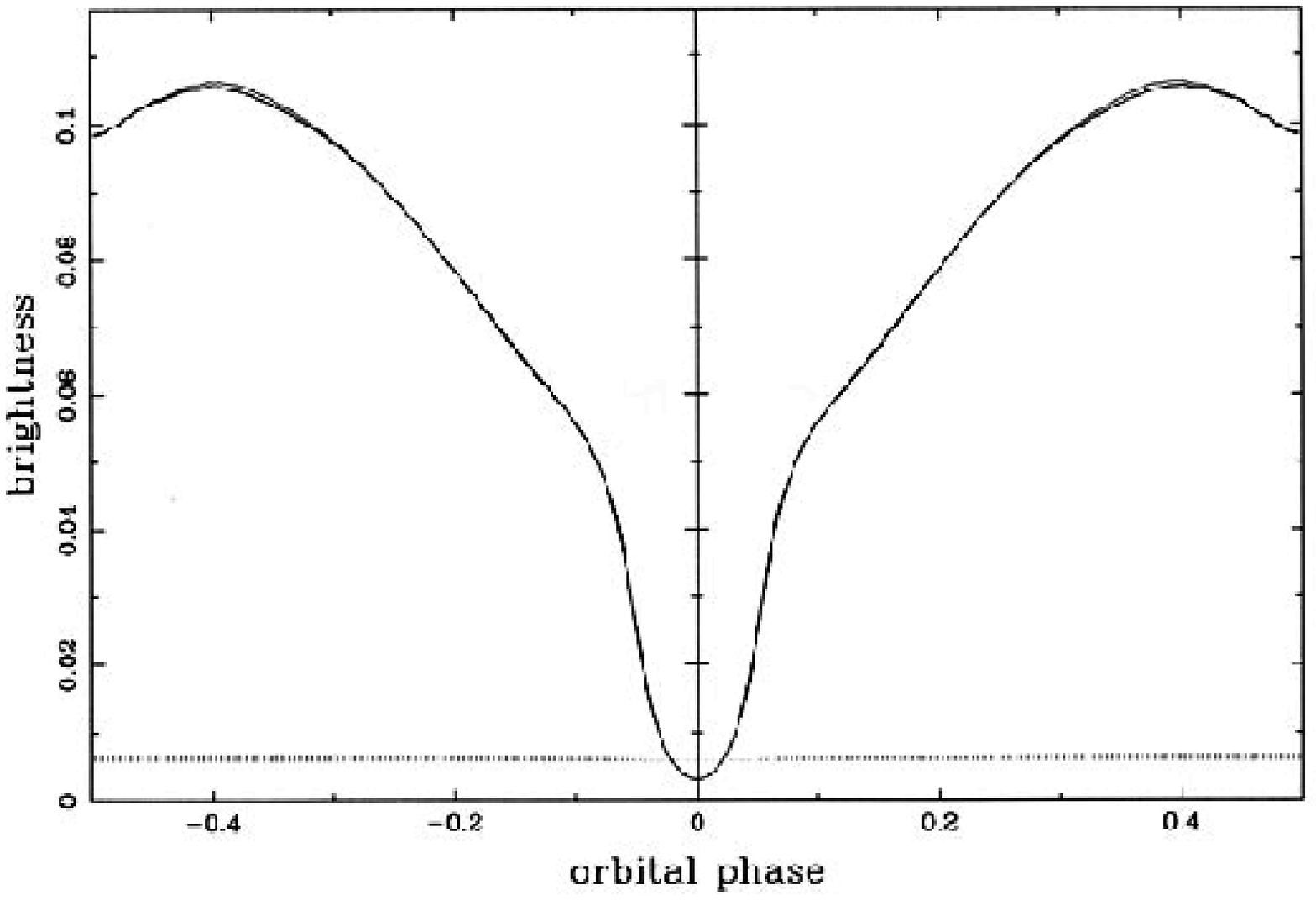}
}
\end{center}
\caption[]{3-D eclipse mapping~\protect\cite{rutten98}.}
\label{fig:rene}
\end{figure}

Wade and Horne~\cite{wade88} did not include an accretion disc grid in
their light-curve fits because the disc does not contribute to the TiO
band-flux. Broad-band light curves, however, especially those obtained
in the infrared, have approximately equal contributions from the disc
and secondary, which means that any light-curve fitting must include
both a secondary star grid and an accretion disc grid.
Rutten~\cite{rutten98} has developed such a technique, known as  {\em
3D eclipse mapping}. The upper-left panel of figure~\ref{fig:rene}
shows a typical accretion disc and secondary star  grid used in 3D
eclipse mapping. A secondary star with a luminous inner hemisphere and
dark outer hemisphere, combined with a standard  $T_{\rm eff}^4
\propto R_{\rm disc}^{-3}$  disc~\cite{warner95}, produces  the light
curve shown in the right-hand panel of figure~\ref{fig:rene}. Fitting
this light curve produces the map of the system shown in the
lower-left panel of figure~\ref{fig:rene}, which is a good
representation of the original intensities assigned to the grid
elements. 3D eclipse mapping has only recently been  applied to real
data~\cite{groot99}, however, and awaits full exploitation in
secondary star studies.

\subsection*{3.1.3~~~Line-width fitting}

We have just seen how variability in two of the `integral' properties
of a  line profile, the line strength and the line position, can be
used to map the secondary stars in CVs. There is, however, a third
integral property of line  profiles which can be used to deduce
surface images -- the variability in line width, or more generally, in
line shape.  The reason line width varies with orbital phase can be
understood by recalling that the secondary stars in CVs are
synchronously rotating and geometrically distorted, which results in a
variation in the projected radius of the secondary as a function of
orbital phase. For a given orbital period, the larger the projected
radius of the secondary the broader the line profile will be, which
means that the rotational broadening varies with orbital phase. At the
conjunction phases, the projected radius and rotational broadening are
at a minimum, while at quadrature they are at a maximum.

The observed variations in rotational broadening also depend on the
inclination of the binary plane and the surface distribution of line
strength. Casares et al.~\cite{casares96}  exploited this effect in
order to determine the inclination of the  magnetic novalike AE~Aqr.
By constructing a secondary star with essentially no surface
structure,  other than that due to an assumed gravity darkening law,
they calculated a grid of model rotational-broadening curves which
were then matched to the observed curve using a $\chi^2$-test.
Shahbaz~\cite{shahbaz98}, also motivated by a desire to determine
inclination angles, added an extra dimension by fitting the line shape
rather than just the line width. Both of these techniques are examples
of {\em model fitting}, in which any surface structure must be added
to the model in an ad-hoc manner prior to fitting. A better approach
to mapping surface structure is {\em image reconstruction}, in which
the  intensity of each image pixel is a free parameter; Roche
tomography is an example of such a technique (see
section~\ref{sec:rochey}).

\subsection{Doppler tomography}

\begin{figure}[t]
\begin{center}
\rotatebox{270}{
\includegraphics[width=1.\textwidth]{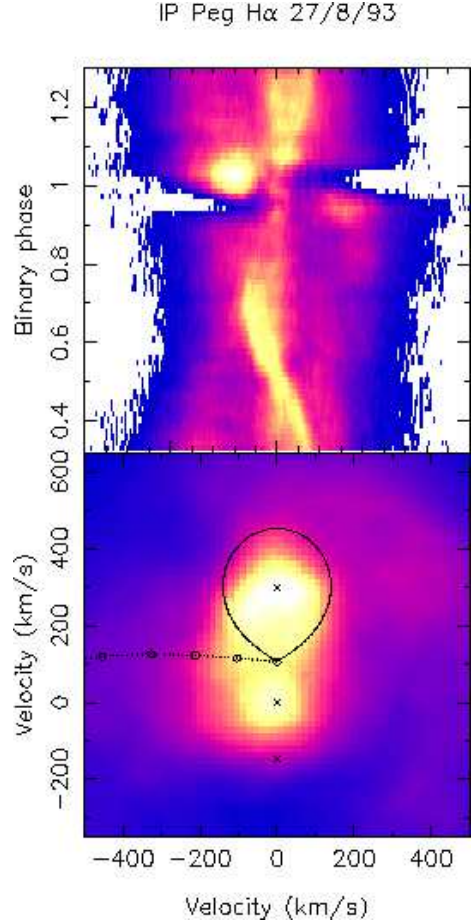}
}
\end{center}
\caption[]{A slingshot prominence and accretion disc shadow in the
dwarf nova IP~Peg~\protect\cite{steeghs96}.}
\label{fig:danny}
\end{figure}

Doppler tomography uses a time-series of emission line profiles
spanning the binary orbit to construct a two-dimensional map of the
system in velocity space. Although primarily used as a tool to study
the  accretion regions in CVs, Doppler tomography has also proved to
be remarkably successful in secondary star studies. This is because
Doppler tomography  provides a way  of cleanly separating the emission
component due to the  secondary star  from the broader and often
stronger emission originating  in the accretion regions. The technique
and some of its most important achievements to date are reviewed
elsewhere in this volume~\cite{marsh00},\cite{schwope00}. In this
section, we highlight one particularly important example of the
application  of Doppler tomography to the study of CV secondaries --
the work of Steeghs  et al.~\cite{steeghs96} on the dwarf nova IP Peg
during  outburst (figure~\ref{fig:danny}).

The upper panel of figure~\ref{fig:danny} shows the observed trailed
spectrum of the H$\alpha$ emission line in IP~Peg. Note the stationary
component at 0 km\,s$^{-1}$ running the entire length of the
time-series and the narrow sinusoidal component which crosses from
red-shifted to blue-shifted around phase 0.5. The sinusoidal emission
component is mapped onto the inner hemisphere of the secondary star's
Roche lobe (lower panel, figure~\ref{fig:danny}).   This emission is
almost certainly being powered by irradiation from the white dwarf and
the hot, inner regions of the outbursting accretion disc, a conclusion
supported by the fact that the emission appears to be stronger around
the poles than it is around the $L_1$ point, implying that obscuration
of the inner disc by the flared, outer disc might be occurring. The
stationary component is mapped onto the centre-of-mass of the binary,
indicated by the cross at zero velocity in the lower panel of 
figure~\ref{fig:danny}. This component is much more difficult to
explain as there is no obvious part of the binary system which is at
rest. One possible interpretation is that the emission is due to
material trapped in a `slingshot prominence', a magnetic loop
originating on and co-rotating with the secondary.  With this model,
Steeghs et al.~\cite{steeghs96} used the Doppler map in
figure~\ref{fig:danny} to estimate the magnetic field strength of  the
secondary star ($\sim$ kG) and the temperature (2 $\times$ 10$^4$ K)
and total mass ($\sim$10$^{18}$ g) of the material in the prominence,
and found good agreement with the parameters deduced from similar
prominences observed in isolated dwarf stars (e.g.~\cite{cameron89}).

\subsection{Roche tomography}
\label{sec:rochey}

Roche tomography~\cite{rutten94b},\cite{watson00} takes as its input a
trailed spectrum, i.e. a time-series of spectral-line profiles, and
provides on output a  map of the secondary star in
three-dimensions. The main advantage of Roche tomography over the
one-dimensional techniques described in section~\ref{sec:1d} is that
the one-dimensional techniques each extract and then use just one
piece of information about the line profile (e.g. the variation in its
flux, radial velocity or width) to map the secondary star, whereas
Roche tomography uses all of the information in the line profile,
as described in section~3.3.1.

Roche tomography is very similar to the Doppler imaging technique used
to map single stars (e.g.~\cite{cameron00},\cite{donati00}), contact
binaries~\cite{maceroni94} and detached secondaries in
pre-CVs~\cite{ramseyer95}. In fact, Roche tomography and Doppler
imaging of single stars differ in only two fundamental ways.  First,
the secondary stars in CVs are tidally-distorted into a Roche-lobe
shape and are in synchronous rotation about the binary centre-of-mass;
isolated stars rotate only about their own centre-of-mass and are
symmetric about their rotation axis.  Second, the continuum is ignored
in Roche tomography, whereas it is included in Doppler imaging. This
is because of  the variable and unknown contribution to the spectrum
of the accretion regions in CVs, forcing Roche tomography to map
absolute line fluxes. The data are therefore slit-loss corrected and
then continuum-subtracted prior to mapping with Roche tomography,
whereas when Doppler imaging the spectra need not be slit-loss
corrected and the continuum is divided into the data.

Although the system parameters are generally better constrained in CVs
than they are in single stars, Roche tomography is a much harder task
than Doppler imaging. This is because CV secondaries are usually both
fainter and more rapidly rotating than isolated stars, resulting in
surface images of a much lower quality than are routinely obtained via
Doppler imaging (see section~3.3.4).

\subsection*{3.3.1~~~Principles and practice}

\begin{figure}[t]
\begin{center}
\rotatebox{270}{
\includegraphics[width=1.\textwidth]{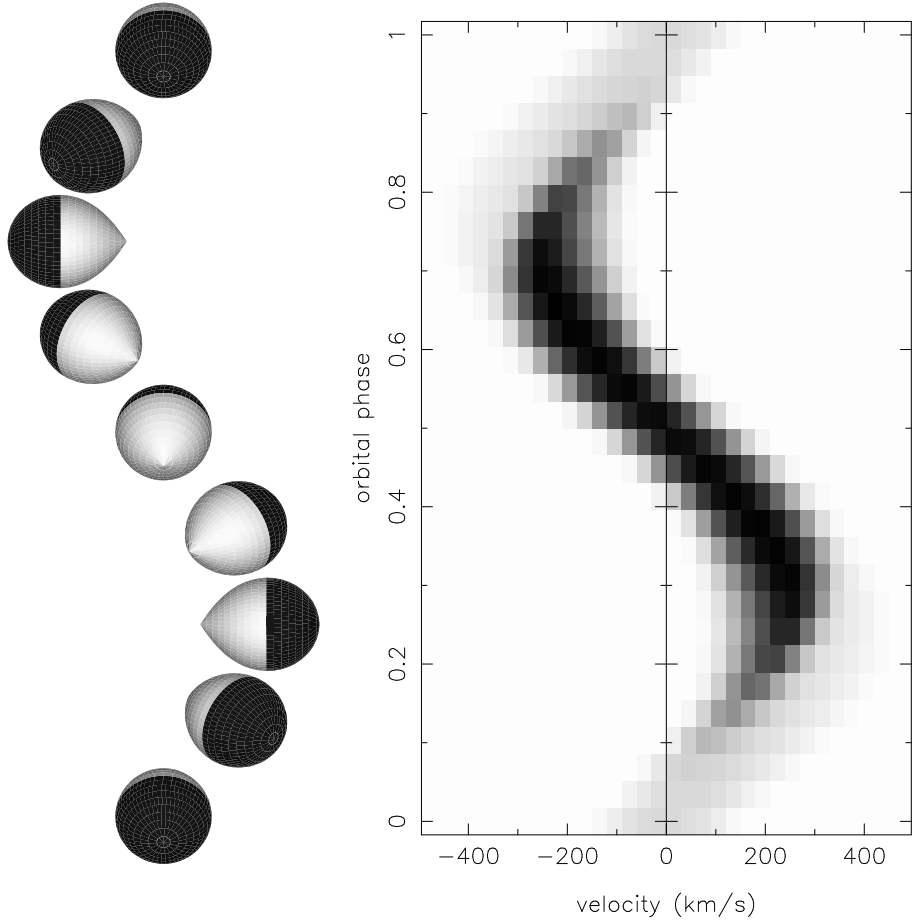}
}
\end{center}
\caption[]{The principles of Roche tomography.}
\label{fig:principles}
\end{figure}

In Roche tomography, the secondary star is modelled as a grid of
quadrilateral surface elements of approximately equal area lying on
the critical potential surface which defines the Roche lobe. Each
surface element, or tile, is then assigned a  copy of the local
specific intensity profile convolved with the instrumental
resolution. These profiles are then scaled to take into account phase
dependent effects, such as variations in the projected area,  limb
darkening and obscuration\footnote{Note that this list does not
include gravity darkening, which is not phase-dependent and hence need
not be included in the algorithm; if any gravity darkening is present
in the data, it will be reconstructed in the maps.} and
Doppler-shifted according to the radial velocity of the surface
element at a particular phase.  Simply summing up the contributions
from each element gives the rotationally broadened profile at any
particular phase. An example of this `forward' process is shown in
figure~\ref{fig:principles}, which shows the trailed spectrum
resulting from a secondary star with a uniformly radiating  inner
hemisphere.

By iteratively varying the strengths of the profile contributed by
each  tile it is possible to perform the `inverse' process and obtain
the map which fits the observed data. How well the map fits the
observed data is  defined by a {\em consistency statistic}, given by
the {\em reduced chi-squared}:
\begin{equation}
{\chi}^2 = {1\over n} \sum_{i=1}^{n} \left( { p_i-o_i} \over {\sigma_i} 
\right)^2,
\label{eq:chiaquared}
\end{equation}
where $n$ is the number of data points, $p_i$ and $o_i$ are the
predicted  and observed data, and $\sigma_i$ is the error on
$o_i$. Fitting the data  as closely as possible (i.e. minimising
$\chi^2$) is not a good approach, however, as noise will dominate the
resulting map. A better approach is to reduce $\chi^2$ until the
observed and predicted data are consistent, i.e. $\chi^{2}\sim 1$.
Such a condition is satisfied by many maps, however, and so a {\em
regularisation statistic} is employed to select just one of
them. Following Horne~\cite{horne85}, we  select the `simplest' map,
which is given by the map of {\em maximum  entropy}. The definition of
entropy, $S$, that we use is
\begin{equation}
S=\sum_{j=1}^{k} m_j-d_j-m_j \ln \left( m_j/d_j \right),
\label{eq:maxent}
\end{equation}
where $k$ is the number of tiles in the map, and $m_j$ and $d_j$ are
the map and the so-called `default map', respectively.
Equation~\ref{eq:maxent} shows that the entropy is  a measure of the
similarity of the map to the default map and hence it is the default
map which defines what we mean by the `simplest' image. The choice of
default map is therefore of great importance. We have experimented
with a  number of different prescriptions for the default
map~\cite{watson00}. The two most successful have been a uniform
default, where every tile in the default is set to the average value
in the map, and a smoothed version of the map, achieved using a
Gaussian blurring  function. In the former case, we are selecting the
{\em most uniform map consistent with the data}, which constrains
large-scale surface structure, and in the latter case we are selecting
the {\em smoothest map consistent with the data}, which constrains
short-scale structure. An efficient algorithm for the task of
maximising entropy subject to the constraint imposed by $\chi^2$ is
given by Skilling and Bryan~\cite{skilling84} and has been implemented
by them in the {\sc fortran} package {\sc memsys}.

\subsection*{3.3.2~~~Assumptions}

The basic assumptions which underlie Roche tomography are:

\begin{enumerate}
\item {\em The secondary star is Roche-lobe filling, synchronously rotating 
and in a circular orbit.}

These are believed to be valid assumptions because there is ample 
evidence for mass transfer and the synchronisation and circularisation 
timescales are both insignificant compared to the lifetime of a 
CV~\cite{warner95}.

\item {\em The observed surface of the star coincides with the surface 
defined by the critical Roche potential.}

The observed surface of a typical CV secondary (i.e. where the
absorption lines originate) lies at a potential corresponding to a
star that fills 99.97 per cent of its Roche lobe, which gives a
change of less than 1 per cent in the shape of the line
profile~\cite{shahbaz98}.

\item {\em The shape of the intrinsic line profile remains unchanged.}

Only the flux of the intrinsic line profile is allowed to vary. 
This assumption is valid because the rotational broadening usually
dominates over intrinsic broadening. Moreover, it should be possible to
identify errors in the intrinsic line-profile shape by the artifacts
they induce in the reconstructions (see section~3.3.3).

\item {\em The line profile is due to secondary star light only.}

The accretion disc is typically too hot to contribute to the cool stellar
features. Furthermore, any accretion disc component present in a
spectral line is usually instantly recognisable due to its extreme
width and anti-phased radial-velocity motion. 

\item {\em The secondary star exhibits no intrinsic variation during the
observation.}

The surface features of the secondary star might change during an observation
due to, for example, a flare. This will result in artifacts in the maps,
as discussed in section~3.3.3.

\item {\em The orbital period, inclination, stellar masses, limb darkening, 
intrinsic line profile and systemic velocity are known.}

The orbital period of a CV is usually precisely known. The other
parameters are more uncertain -- we explore how errors in them affect
the reconstructions in section~3.3.3.

\item {\em The final map is the one of maximum entropy (relative to an 
assumed default image) which is consistent with the data.}

The data constrains the final map through the consistency statistic,
$\chi^2$. The default map constrains the final map through the
regularisation statistic, $S$. If the data are noisy, the data
constraints will be weak and the map will be strongly influenced by
the default map.  If the data are good, however, the image will not be
greatly influenced by the default and the choice of default makes
little difference to the  final map. The default may thus be regarded
as containing prior information about the map, and the map will be
modified only if the observations require it. The importance of this
assumption therefore depends on the quality of the data.
\end{enumerate}

\subsection*{3.3.3~~~Errors}

The maps resulting from any form of astro-tomography are prone to both
systematic errors, due to errors in the assumptions underlying the
technique, and statistical errors, due to measurement errors on the
observed data  points. It is essential that the effects of these
errors on the reconstructions are quantified in order to properly
assess the reality of any surface structure present. This is
especially true of Roche tomography, for which the input data is
generally noisier due to the faintness of CV secondaries, a problem
further exacerbated by the fact that noise in Roche tomograms can
mimic the appearance of star-spots~\cite{watson00}.

\begin{figure}[t]
\rotatebox{270}{\includegraphics[width=0.315\textwidth]{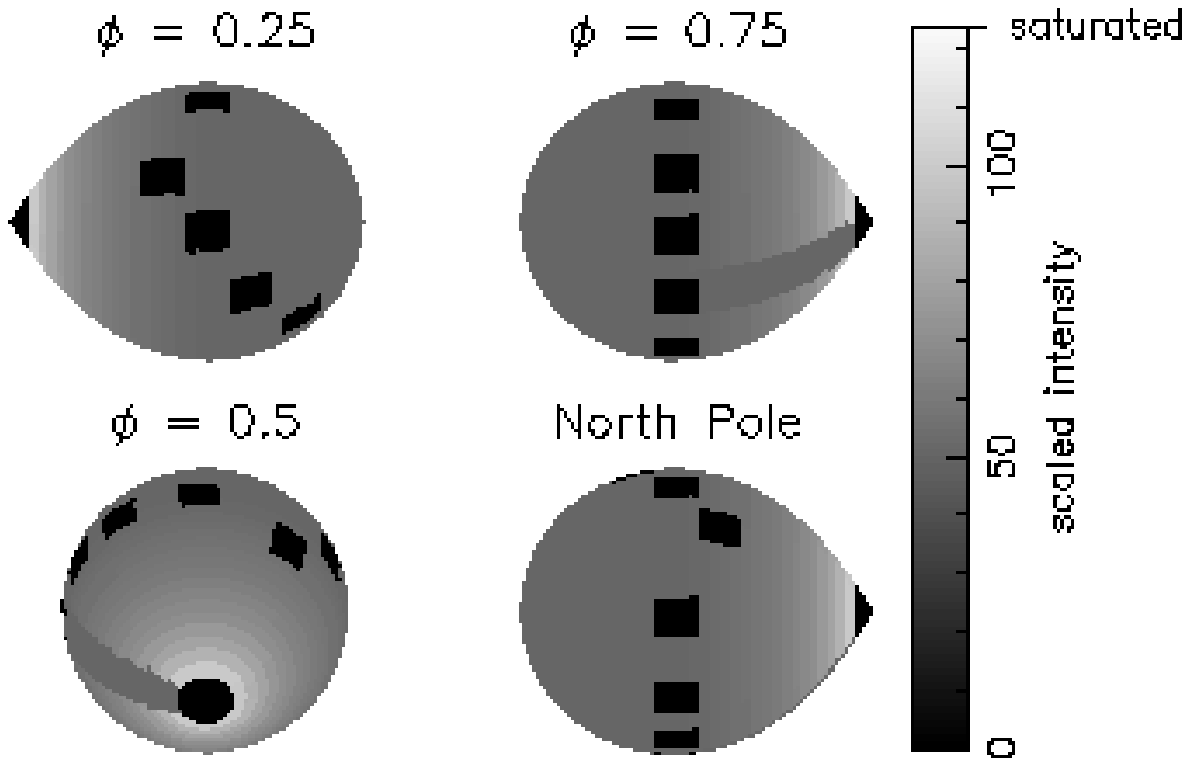}}~\rotatebox{270}{\includegraphics[width=0.315\textwidth]{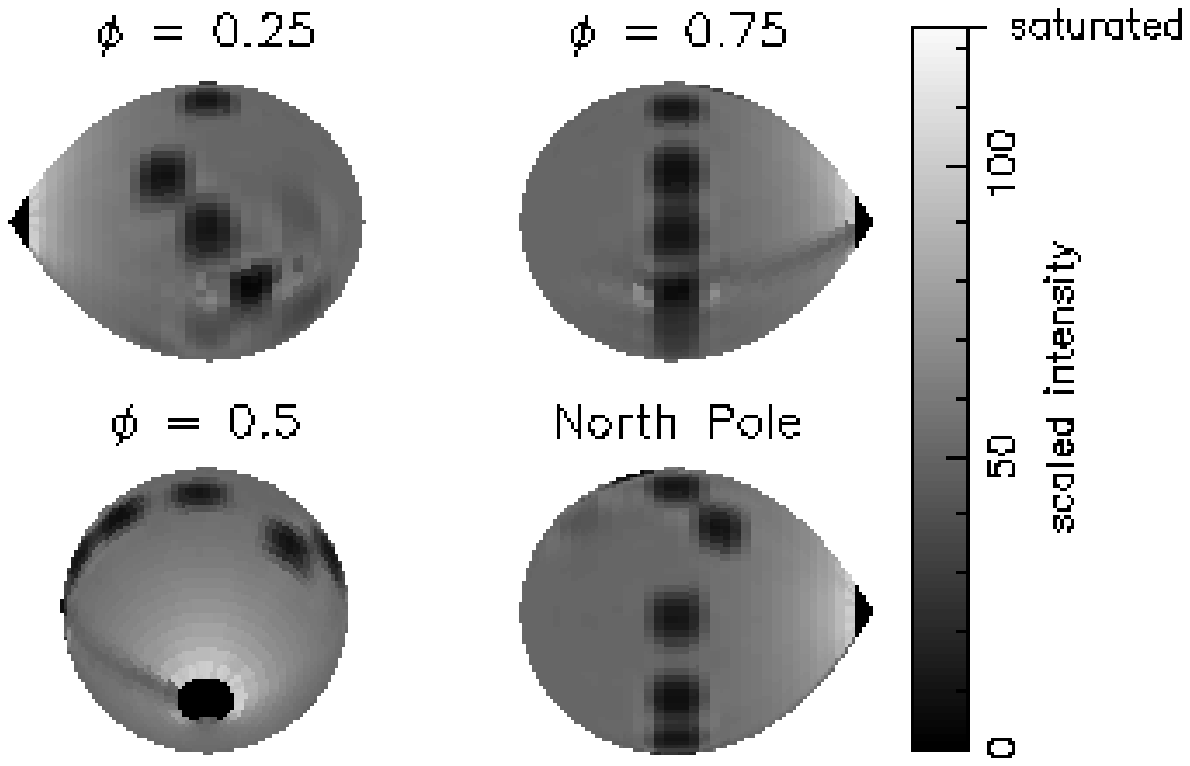}}

\vspace*{0.5cm}

\rotatebox{270}{\includegraphics[width=0.315\textwidth]{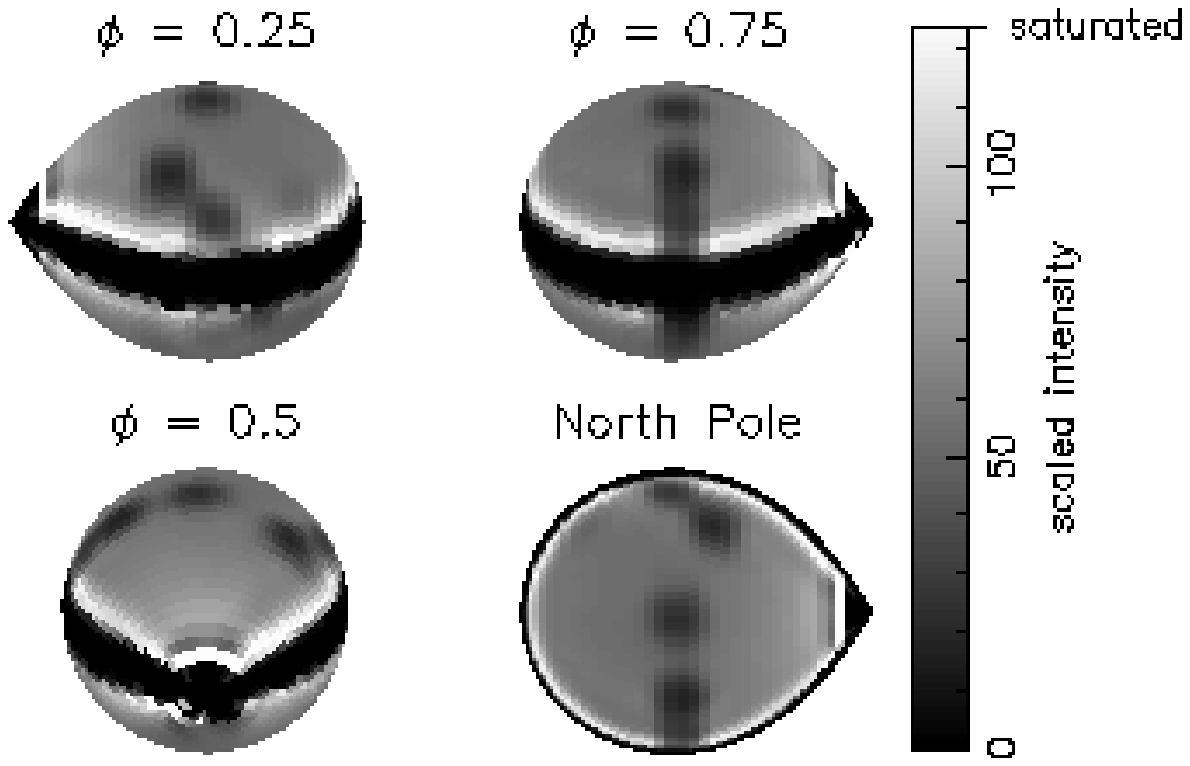}}~\rotatebox{270}{\includegraphics[width=0.315\textwidth]{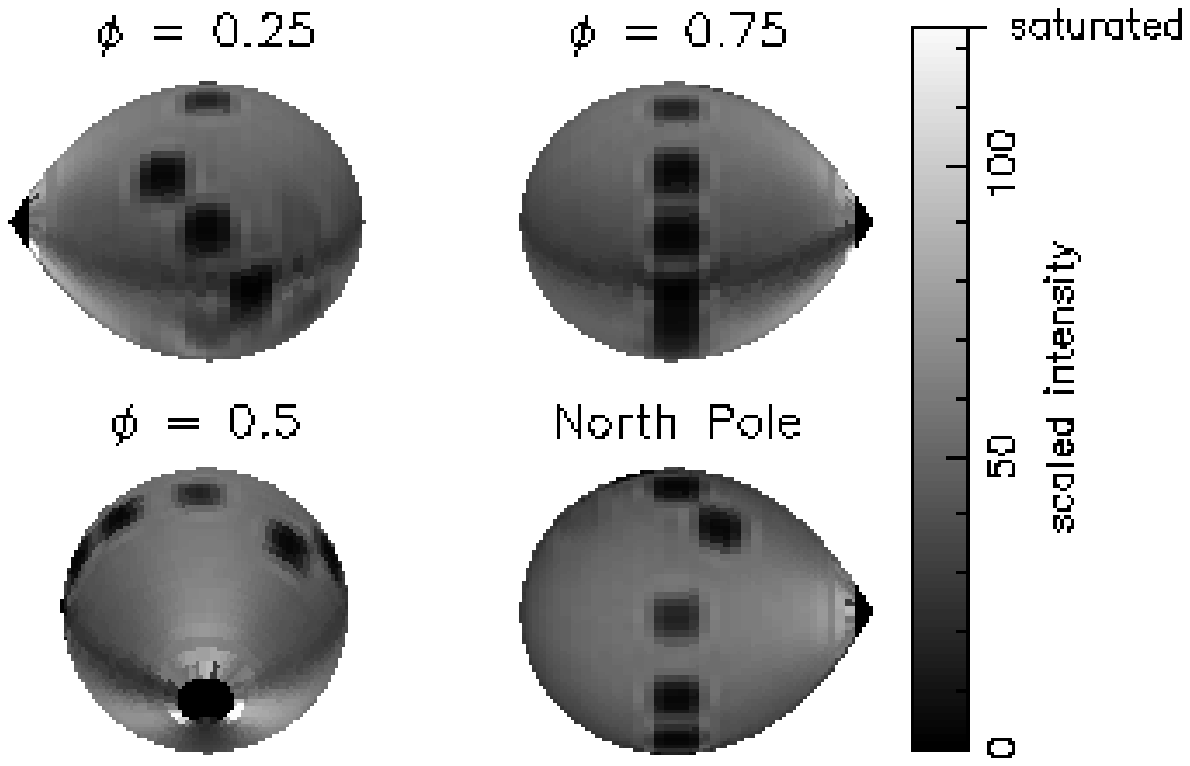}}

\vspace*{0.5cm}


\vspace*{0.5cm}

\rotatebox{270}{\includegraphics[width=0.315\textwidth]{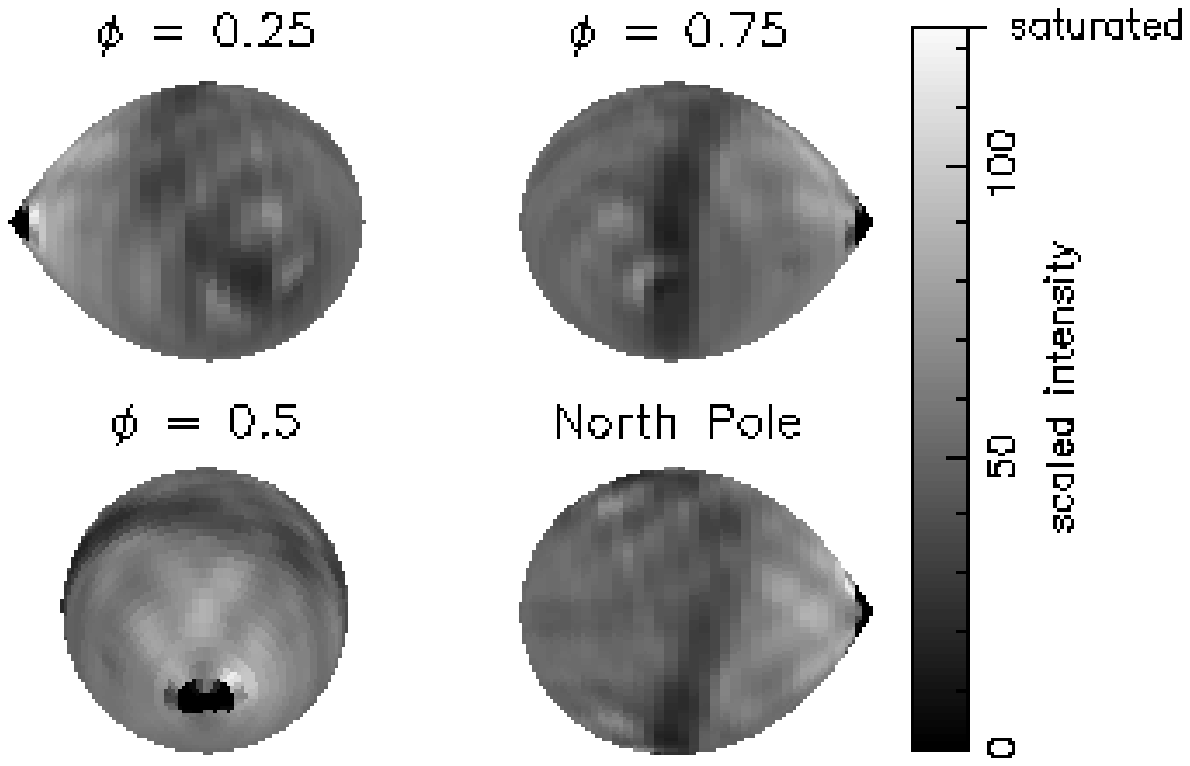}}~~~\rotatebox{270}{\includegraphics[width=0.315\textwidth]{artefacts.ps}}
\caption[]{Systematic errors in Roche tomography~\cite{watson00}.  Top
row: The test image (left) and the best fit (right).  Middle row:
Equatorial banding due to the effect of an incorrect systemic velocity
(left) and incorrect limb darkening (right). Bottom row: Ring-like
streaks due to the  effects of phase undersampling (left), which
correspond to  lines on the secondary star with the same radial
velocities as the spots (right).}

\label{fig:systematics}
\end{figure}

We begin by discussing our approach to statistical error
determination~\cite{watson00}.  The maximum-entropy technique is
non-linear, in the sense that each map value is not a linear function
of the data values. This makes it very difficult to propagate the
statistical error on a data point through the maximum-entropy process
in order to calculate the statistical error on each tile of the
map. Furthermore, the statistical errors on each tile  will not be
independent due to, for example, the projection of bumps in the line
profiles across arcs of constant radial velocity on the secondary star
(see figure~\ref{fig:systematics}). The simplest approach to error
estimation, in this case, is to use a Monte Carlo-based simulation.

Monte-Carlo techniques rely on the construction of  a large (typically
hundreds, in the case of Roche tomography) sample of synthesized
datasets  which have been effectively drawn from the same parent
population as the  original dataset, i.e. as if the observations have
been repeated many  hundreds of times. This large sample of
synthesized datasets is then used to  create a large sample of Roche
tomograms, resulting in a probability distribution for each tile in
the map.  The main difficulty with this technique, aside from the
demands on  computer time, lies in the construction of the sample of
synthesized datasets. One approach (e.g.~\cite{rutten94a}) is  to
`jiggle' each data point about its observed value, by an
amount given by its error bar multiplied by a number output by a
Gaussian random-number generator with zero mean and unit
variance. This process adds noise to the data, however, which means
that the synthesized datasets are not being drawn from the same parent
population as the observed dataset -- noise is being added to a
dataset which has already had noise added to it during the measurement
process. In practice, this means that fitting the sample datasets to
the same level of $\chi^2$ as the original dataset is either
impossible (i.e. the iteration does not converge) or results in maps
dominated by noise, which overestimates the true error on each tile.

A much better approach to creating synthesized datasets is the {\em
bootstrap method}~\cite{efron82}, which we have implemented as
follows~\cite{watson00}.  From our observed trailed spectrum
containing $n$ data points, we select, {\em at random and with
replacement},  $n$ data values and place them in their original
positions in the new,  synthesized trailed spectrum. Some points in
the synthesized trailed spectrum will be empty, in which case they
will be omitted from the fit; in practice, this is achieved by setting
the error bars on these points to infinity. Other points in the
synthesized  trailed spectrum will have been selected once or more, in
which case  their error bars are divided by the square root of the
number of times  they were picked. The advantage of bootstrap
resampling over jiggling is  that the data is not made noisier  by the
process as only the errors bars on the data points are manipulated. It
is therefore possible to fit the synthesized trailed spectra to the
same  level of $\chi^2$ as the observed data, giving a much more
reliable estimate of the statistical errors in the maps.  The
bootstrap method has been shown to work very effectively with Roche
tomography~\cite{watson00}, and we present an example of its
application in
figures~\ref{fig:huaqr_errors}~and~\ref{fig:ippeg_errors}.

The determination of systematic errors requires a completely different
approach to that employed for the determination of statistical errors.
We have approached systematic errors in Roche
tomography~\cite{watson00}  in much the same way that Marsh and
Horne~\cite{marsh88a} explored them in Doppler tomography. We first
constructed a test image (top-row, left, figure~\ref{fig:systematics})
of the secondary star containing all of the surface features we might
expect to find when mapping real data. These include irradiation,
shadowing by the gas stream and star-spots covering a range of
latitudes and longitudes. We used the test image to create a  test
trailed spectrum which we then fit using Roche tomography to
reconstruct the surface image of the star. As can be seen from
figure~\ref{fig:systematics} (top-row, right), the best fit reproduces
all  of the features in the test image. We then explored the effects
of systematic errors on the reconstruction by varying the input
parameters to Roche tomography and re-fitting the test trailed
spectrum. In this way we explored how variations in the  systemic
velocity, limb darkening, inclination, velocity smearing, phase
undersampling, noise, intrinsic line profile, resolution, default map,
stellar  masses and stellar flares affect the
reconstructions~\cite{watson00}.

We find that systematic errors in Roche tomography generally result in
only two major artifacts in the Roche tomograms: ring-like streaks and
equatorial banding~\cite{watson00}.  Equatorial banding results
whenever there is a mis-match between the  maximum velocities  present
in the line  profile and the maximum velocities available on the
secondary star grid, which  occurs whenever the wrong systemic
velocity, limb darkening, inclination, velocity smearing, intrinsic
line profile or stellar masses are used in the reconstruction. The
banding occurs around the equator of the star because this is where
the highest radial velocities  are found. Some examples of equatorial
banding  patterns in Roche tomograms are shown in
figure~\ref{fig:systematics},  where the maps have been reconstructed
using incorrect values for the  systemic velocity (middle row, left)
and limb darkening (middle row, right).

Ring-like streaks appear in Roche tomograms when one or a few  of the
spectra in the input data dominate. This occurs when there is phase
undersampling, for example, or when there is a flare in the data at a
particular phase. The effect is analogous to the streaks observed in
Doppler tomography~\cite{marsh88a} and can be understood by
considering that lines of constant radial velocity on the secondary
star at a particular phase can be integrated along to construct a line
profile.  These lines of constant radial velocity are ring-like in
shape and if there are only a few phases, or if the profile is
particularly bright at a certain phase, the streaks will not
destructively interfere, leaving ring-like artifacts on the Roche
tomogram. An example of a Roche tomogram reconstructed using only 5
orbital phases is shown in figure~\ref{fig:systematics} (bottom row,
left). The streaks present in this image correspond to lines on the
secondary star with the same radial velocity as the spots, as shown at
phase 0.25 in figure~\ref{fig:systematics} (bottom row, right).

The error experiments presented here show that any feature on a Roche
tomogram must be subjected to two tests before its reality can be
confirmed. The first test is to determine whether the feature is
statistically significant and is performed via a  Monte-Carlo
technique with bootstrap-resampling. The second test is to compare the
feature with the appearance of known artifacts of the technique due to
errors in the underlying assumptions, such as those presented in
figure~\ref{fig:systematics}.  If a surface feature survives both of
these tests unscathed, it can be assumed to be real.

\subsection*{3.3.4~~~Results}

Roche tomograms currently exist for only 3 CVs: the novalike
DW~UMa~\cite{rutten94b}, the dwarf nova IP~Peg~\cite{rutten96} and the
polar AM~Her~\cite{davey96}. In this section we will present two new
Roche tomograms -- the polar HU~Aqr and an improved study of IP~Peg.

\begin{figure}[t]
\begin{center}
\rotatebox{270}{
\includegraphics[width=0.745\textwidth]{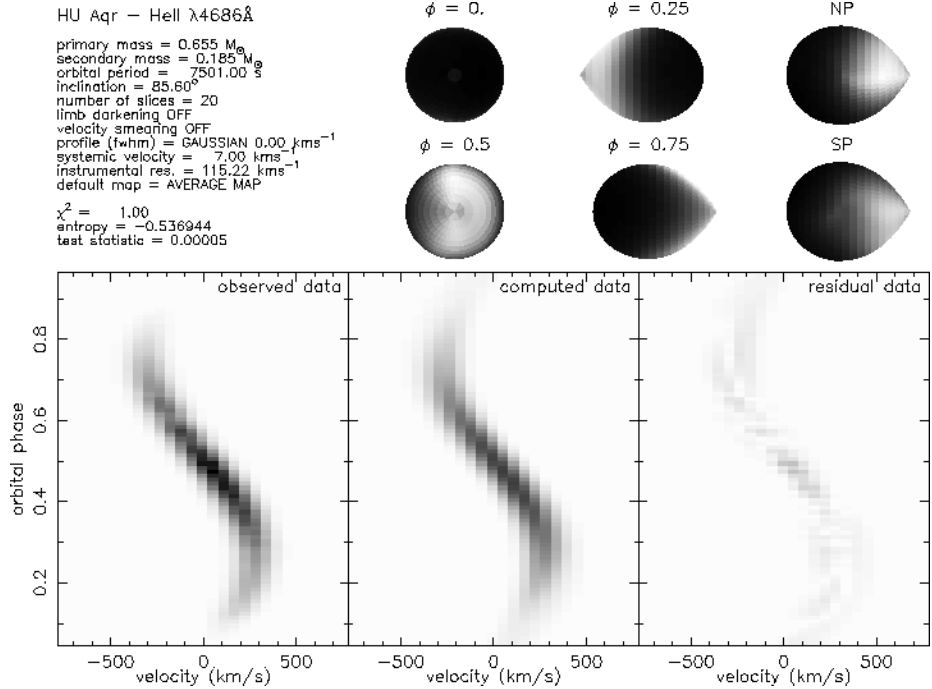}
}
\end{center}
\caption[]{Roche tomogram of the polar HU~Aqr.}
\label{fig:huaqr_rdisp}
\end{figure}

\begin{figure}[t]
\begin{center}
\rotatebox{270}{
\includegraphics[width=0.725\textwidth]{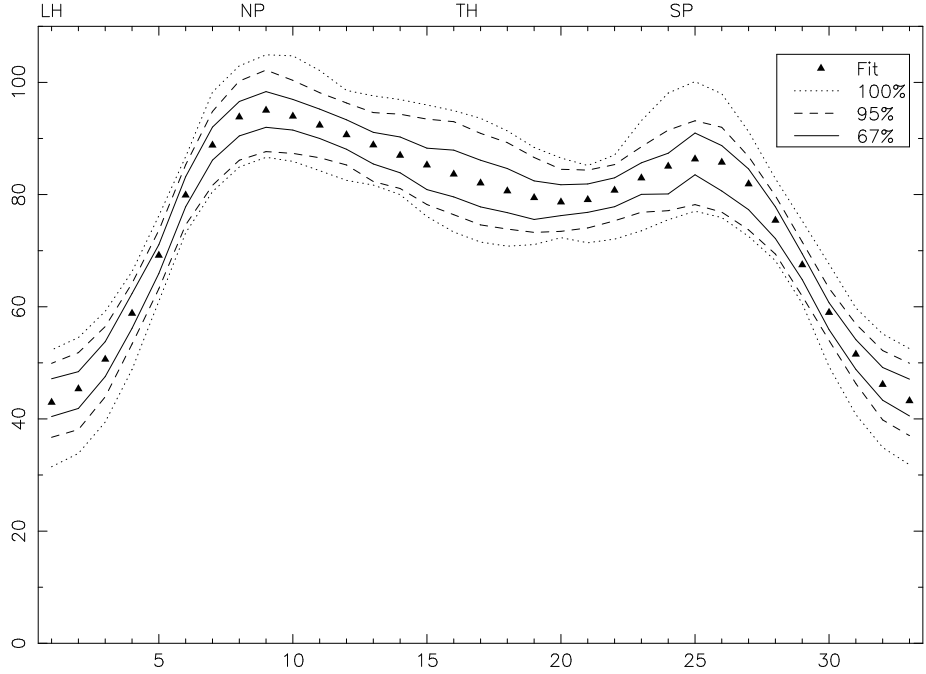}
}
\end{center}
\caption[]{A slice through the secondary star in HU~Aqr.} 
\label{fig:huaqr_errors}
\end{figure}

The Roche tomogram of HU~Aqr in the light of the He\,II $4686$\AA\
emission line observed by Schwope et al.~\cite{schwope97} is shown in
figure~\ref{fig:huaqr_rdisp}, where bright regions in the map
represent areas where the emission line flux is at its
strongest. There are two main features in the Roche tomogram -- the
strong asymmetry between the inner (phase 0.5) and outer (phase 0)
hemispheres of the secondary  and a weaker asymmetry between the
leading (phase 0.75) and trailing (phase 0.25) hemispheres of the
secondary. The  reality of the latter asymmetry can be assessed from
figure~\ref{fig:huaqr_errors}, which shows a slice passing through the
leading hemisphere (LH), north pole (NP), trailing hemisphere (TH)and
south pole (SP) of the secondary star.  The triangular points
represent the map values, and it can be seen that the leading
hemisphere is a factor of two fainter than the trailing
hemisphere. The significance of this difference can be  assessed from
the curves in figure~\ref{fig:huaqr_errors}, which show confidence
limits on the map values derived from 200 bootstrap resampling
experiments; 67 per cent of the map values (measured relative to the
mode of the distribution)  lie within the range bounded by the  solid
curves and 100 per cent of the map values lie within the range
bounded by the dotted curves.  Figure~\ref{fig:huaqr_errors} shows
that we can be 100 per cent certain that the asymmetries between the
trailing and leading hemispheres are not due to statistical errors.
Furthermore, none of the systematic errors explored in section~3.3.3
result in asymmetries of the type observed in
figure~\ref{fig:huaqr_rdisp}, implying that the asymmetries are
real. Schwope et al.~\cite{schwope97} reached a similar conclusion
based on a model-fitting technique. They attributed the asymmetry
between the inner and outer hemispheres to irradiation by the magnetic
accretion column and the asymmetry between the leading and trailing
hemispheres to shielding of this irradiation by the gas stream.

\begin{figure}[t]
\begin{center}
\rotatebox{270}{
\includegraphics[width=0.745\textwidth]{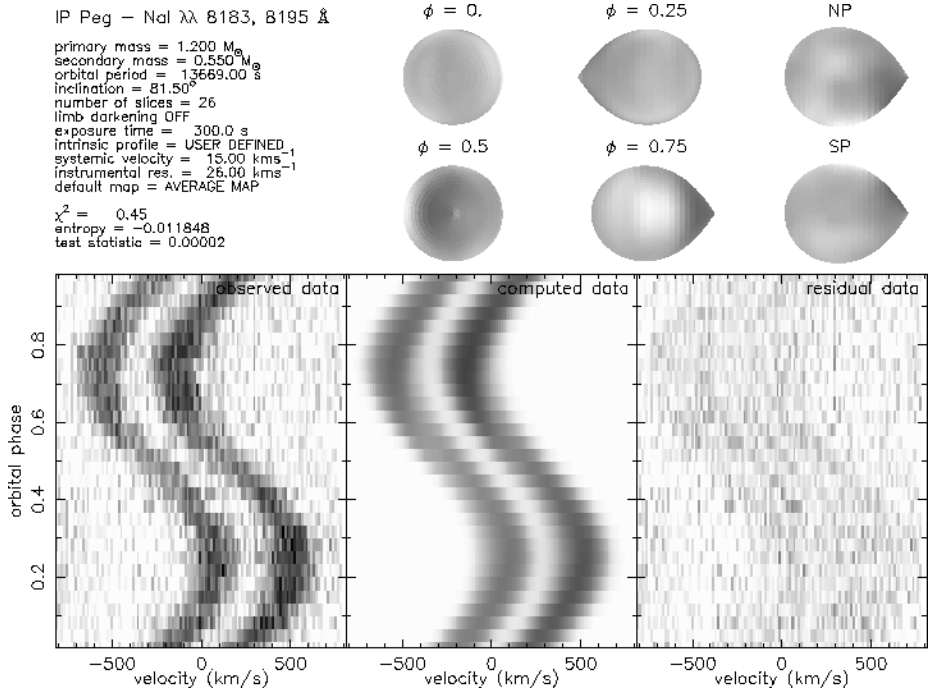}
}
\end{center}
\caption[]{Roche tomogram of the dwarf nova IP~Peg.}
\label{fig:ippeg_rdisp}
\end{figure}

\begin{figure}[t]
\begin{center}
\rotatebox{270}{
\includegraphics[width=0.725\textwidth]{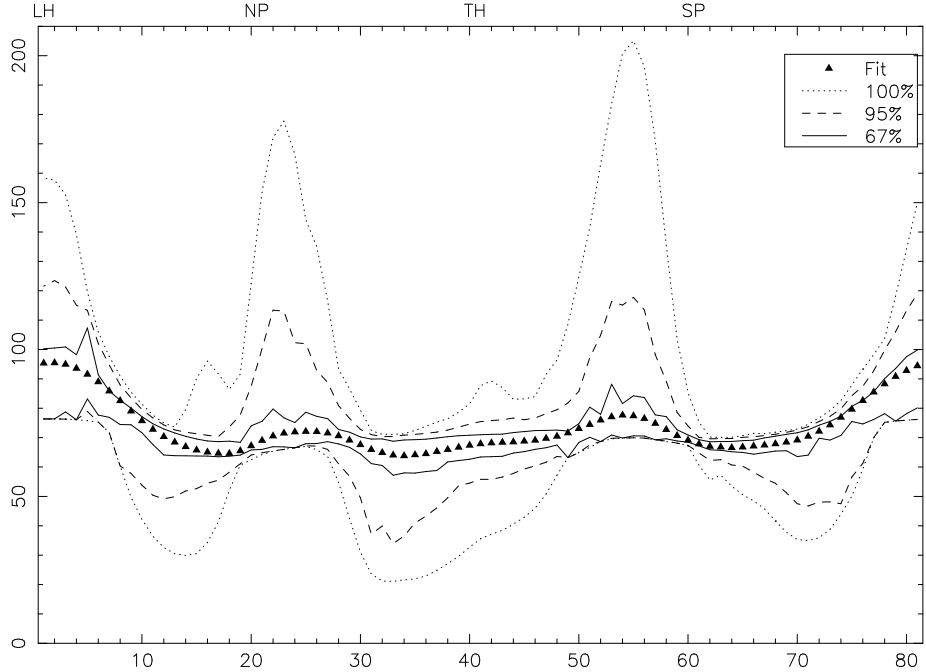}
}
\end{center}
\caption[]{A slice through the secondary star in IP~Peg.} 
\label{fig:ippeg_errors}
\end{figure}

The Roche tomogram of IP~Peg in the light of the Na\,I absorption
doublet around 8190\AA\ is presented in figure~\ref{fig:ippeg_rdisp},
where bright regions in the map represent areas where the flux deficit
is at its largest, i.e. where the absorption line is at its strongest.
The most noticeable feature in the map is the asymmetry between the
inner (phase 0.5) and outer (phase 0) hemispheres, which is slightly
skewed towards the leading edge (phase 0.75) of the secondary. As
described in section~3.1.1, this can be attributed to the effect of
irradiation by the accretion regions and the resulting circulation
currents on the secondary. There is another feature of note in the
tomogram of figure~\ref{fig:ippeg_rdisp} -- a spot on the leading edge
of the secondary. The spot is bright, which means it is an area that
exhibits stronger Na\,I absorption. This is not consistent with what
one might expect from a star-spot, because it is known that the Na\,I
flux deficit decreases with later spectral type~\cite{brett93} and
star-spots are generally believed to be of order 1000~K cooler than
the surrounding photosphere~\cite{vogt81} (although spots hotter than
the photosphere have also been imaged~\cite{donati92}). An inspection
of the confidence limits in  a slice through the spot
(figure~\ref{fig:ippeg_errors}), however, shows that the feature is
not statistically significant; the spot is  the hump at `LH' in
figure~\ref{fig:ippeg_errors} and it can be seen that the 67 per cent
confidence limits widen significantly around it.

\begin{figure}[t]
\begin{center}
\rotatebox{270}{
\includegraphics[width=0.77\textwidth]{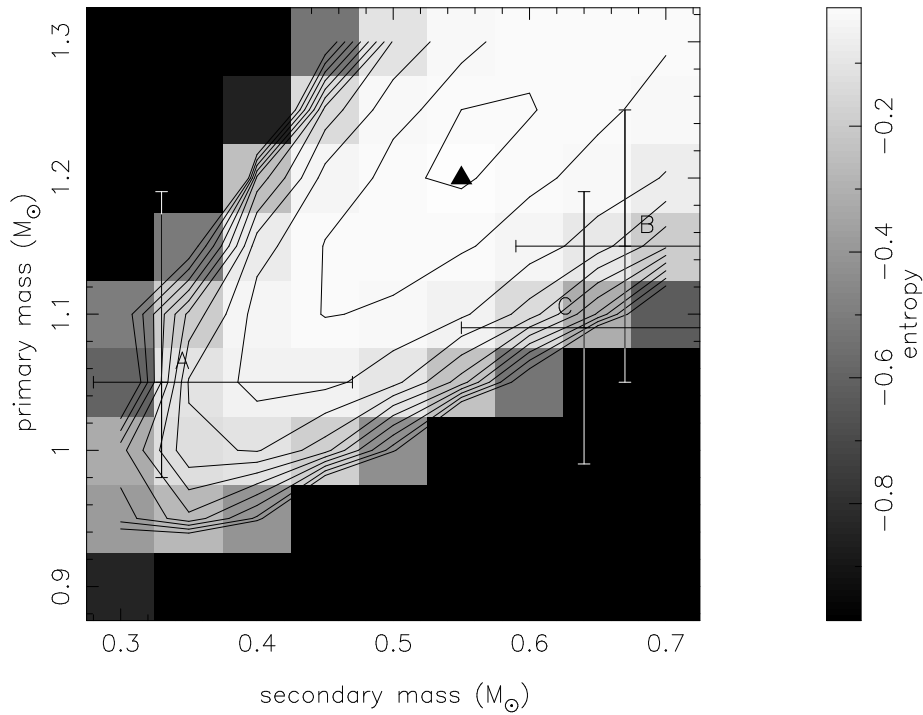}
}
\end{center}
\caption[]{Entropy landscape for IP~Peg.}
\label{fig:ippeg_land}
\end{figure}

Although we have not, unfortunately, imaged a star-spot in IP~Peg, we
can use the Roche tomogram to derive accurate values for the masses of
the stellar components. As discussed in section~3.3.3,  using
incorrect values for stellar masses results in equatorial banding,
which increases the entropy of the reconstructions when using a
uniform default map. The masses (and inclinations consistent with the
masses and the observed eclipse width) can therefore effectively  be
included in the fit by constructing an {\em entropy
landscape}~\cite{rutten94b},\cite{rutten96}, which plots entropy as a
function of the primary and secondary masses. The entropy landscape
for IP~Peg is shown in figure~\ref{fig:ippeg_land}. The triangle
denotes the point of highest entropy, corresponding to primary and
secondary masses of $M_1=1.2$ M$_\odot$ and $M_2=0.55$ M$_\odot$.
These values are in approximate agreement with the mass determinations
of Beekman et al.~\cite{beekman00}, Martin et al.~\cite{martin89}  and
Marsh~\cite{marsh88b}, marked $A$, $B$ and $C$ in
figure~\ref{fig:ippeg_land}.  Because the masses derived from an
entropy landscape automatically  account for the geometrical
distortion of the secondary and any  non-uniformities in its  surface
structure, the technique provides  very tightly constrained mass
determinations, particularly in directions  orthogonal to the diagonal
feature in figure~\ref{fig:ippeg_land} (corresponding to mass ratios
that allow the radial velocity of the  secondary's centre-of-mass to
remain  constant).

\begin{figure}[t]
\begin{center}
\rotatebox{270}{
\includegraphics[width=0.725\textwidth]{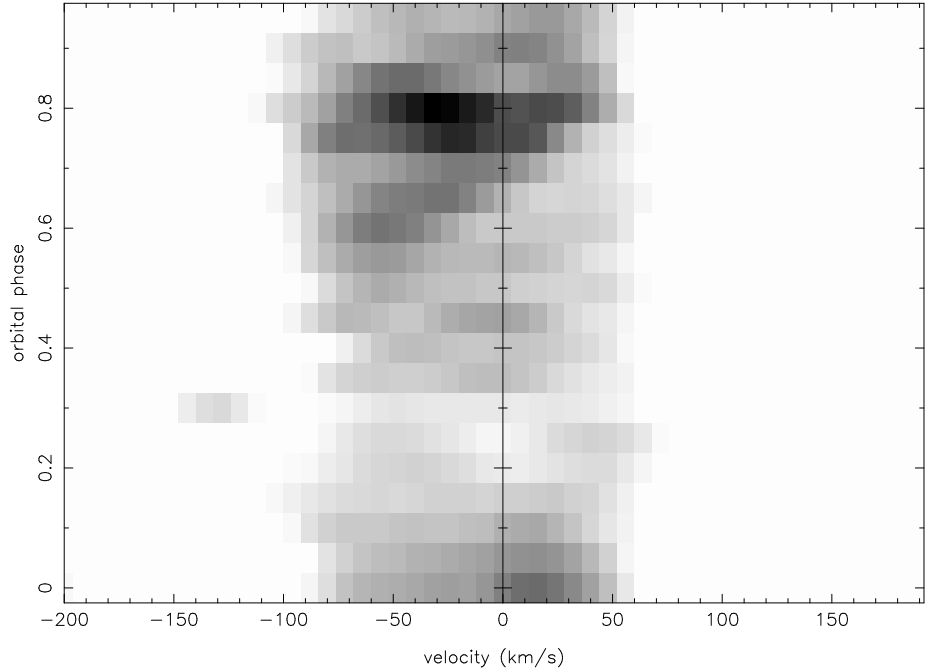}
}
\end{center}
\caption[]{Trailed spectrum of SS Cyg derived using 
LSD~\protect\cite{shahbaz00}.}
\label{fig:sscyg}
\end{figure}

\section{Conclusions}

The Roche tomogram of IP~Peg presented in figure~\ref{fig:ippeg_rdisp}
represents the best that can be obtained with a 4-m class telescope
when mapping a single spectral line. Yet even this does not appear to
be good  enough to image the star-spots we might reasonably expect to
be present on IP~Peg's secondary. To provide routine imaging of
star-spots on CV secondaries it will therefore be necessary to move up
to 8-m class telescopes and/or combine many spectral lines to increase
signal-to-noise, using techniques such as {\em least-squares
deconvolution}  (LSD)~\cite{donati97}. An initial attempt at LSD using
data on the dwarf nova SS~Cyg is presented in figure~\ref{fig:sscyg}
and appears to show  the tell-tale signature of a star-spot -- the
diagonal stripe moving from  blue to red velocities between phases
0.5--0.8. These data require more careful reduction before they can be
mapped, but it is clear that LSD is the way forward and we can expect
the first star-spots on CV secondaries to have been unambiguously
imaged  by the time the next conference on astro-tomography convenes.

\section*{Acknowledgments}
We thank Andrew Cameron, Jean-Fan\c{c}ois Donati, Tom Marsh, Ren\'{e}
Rutten, Axel Schwope, Tariq Shahbaz and Danny Steeghs for allowing us
to present their data in this review and for much useful  advice on
the art of astro-tomography.

\end{document}